\documentclass[
superscriptaddress,
twocolumn,
amsmath,amssymb,
aps,
prl,
]{revtex4-1}

\usepackage{graphicx} 
\usepackage{dcolumn} 
\usepackage{bm} 
\usepackage{braket}
\usepackage{color}
\usepackage{amsthm}

\usepackage{tikz}

\usepackage{hyperref} 
\usepackage{qcircuit}       

\definecolor{lightblue}{RGB}{73,151,208}
\definecolor{crimson}{RGB}{140,41,53}

\hypersetup{
    colorlinks,
    linkcolor={crimson},
    citecolor={lightblue},
    urlcolor={lightblue}
}

\def\>{\rangle}
\def\<{\langle}

\usepackage{mathtools}

\DeclarePairedDelimiter{\ceil}{\lceil}{\rceil}

\begin{document}

\preprint{APS/123-QED}

\title{Generalized Parity Measurements and Efficient Large Multi-component Cat State Preparation with Quantum Signal Processing} 

\author{Sina Zeytino\u glu }
\affiliation{
Physics and Informatics Laboratory, NTT Research, Inc.,940 Stewart Dr., Sunnyvale, California, 94085, USA }

\begin{abstract}
 
Generalized parity measurements are instrumental for the preparation of non-trivial quantum states and the detection of errors in error correction codes. Here, we detail a proposal for efficient and robust generalized parity measurements based on Quantum Signal Processing. Most strikingly, given access to an evolution generated by a one-to-all coupling interaction Hamiltonian between a measurement qubit and the measured system, the desired measurement can be implemented in constant time determined only by the interaction rate. 
The proposed generalized parity measurement can be used to efficiently prepare high-fidelity multi-component cat states in the setting of superconducting cavity quantum electrodynamics.
We benchmark the state-preparation protocol through numerical simulations with realistic system parameters. We show that a 20-component cat state with $400$ photons can be prepared with success probability $>2\%$ and a fidelity $\approx 90\%$ limited by the cavity decay and nonlinear qubit-cavity coupling rates.
Our results pave the way for the realization of a wide range of useful non-classical states consisting of a large number of excitations.
\end{abstract}
\date{\today}%

\maketitle

Measurements are of great importance for quantum technologies. On the one hand, they can be used to prepare entangled quantum states that are heralded by the result of the measurement \cite{ionicioiu2008generalized,childs2000finding, ruskov2003entanglement,tantivasadakarn2024long}. On the other hand, they allow experimentalists to act on a quantum system conditioned on particular measurement outcomes. This feedback loop is at the heart of all efforts to control quantum mechanical systems \cite{lloyd2000coherent,touchette2000information} enabling all levels of quantum control from system identification \cite{gebhart2023learning} to quantum error correction \cite{lloyd1993potentially,shor1995scheme}, and universal information processing paradigms such as measurement-based quantum computation \cite{knill2001scheme,kok2007linear,raussendorf2001one,briegel2009measurement}.  
In the context of measurements on quantum systems with many degrees of freedom,
parity measurements, which project the measured system onto subspaces with even or odd number of excitations \cite{ruskov2003entanglement} proved to be particularly important. Parity measurements have been discussed in the context of graph state preparation \cite{pittman2001probabilistic} for measurement-based quantum computation \cite{gottesman1999demonstrating,pittman2001probabilistic,Rebentrost2014}, as well as for error detection in a wide range of quantum error correction protocols in spin and bosonic systems \cite{kitaev1995quantum,devitt2013quantum,terhal2015quantum,sun2014tracking}.

Recent years have also witnessed a growing interest in generalized parity measurements
(hereafter denoted ${\rm GP}_{r,k}$), whose binary outcome 
projects the system into or out of a subspace $S_{k,r}$ spanned by states $\ket{m}$ with $m = k\mod{r}$ excitations.
Formally, the generalized parity measurement ${\rm GP}_{r,k}$ has a binary outcome, 
described by the measurement operators $M^{r,k}_{0}$ and $M^{r,k}_{1}$ 
that project $\mathcal{S}$ onto $S_{r,k}$ and its complement, respectively.
Such generalized parity measurements were first introduced for spin systems in Ref. 
\cite{ionicioiu2008generalized}. There, the authors showed how ${\rm GP}_{r,k}$ can be used to prepare entangled states such as the $n$-qubit GHZ and W states, and other Dicke states, which are of interest for quantum communication applications \cite{ionicioiu2008generalized,prevedel2009experimental}, metrology \cite{toth2012multipartite,pezze2018quantum,hakoshima2020efficient,wang2021preparing} and variational quantum optimization algorithms \cite{bartschi2019deterministic,barnes2022assembly}. Generalized parity measurements were brought back into focus by new and exciting applications in error-correcting codes encoded in bosonic  \cite{michael2016new,albert2018performance,ni2023beating,jain2024quantum} and qubit systems \cite{omanakuttan2024fault}, as well as applications in photon-number resolved measurements \cite{curtis2021single}. 

Despite the promising use cases, the realization of generalized parity measurements on near-term platforms is hampered by the difficulty of their implementation, especially compared to conventional parity measurements (i.e., $r=2$). 
Indeed, it is relatively straightforward to design a protocol to implement conventional parity measurements 
using a single ancillary (measurement) qubit that is coupled to all system qubits \cite{kok2007linear,krinner2022realizing,google2023suppressing,riste2013deterministic} or to a bosonic degree of freedom \cite{Gambetta2006qubit,blais2020quantum}.  
However, such protocols 
cannot be directly extended to generalized parity measurements without increasing the size of the ancillary measurement system \cite{ionicioiu2008generalized} or the number of control channels on a single qubit, as in the frequency multiplexed measurement in Refs. \cite{michael2016new,essig2021multiplexed,Gambetta2006qubit,schuster2007resolving}.

Recently, an optimal control approach has been applied to the problem of designing generalized parity measurement protocols for bosonic systems \cite{curtis2021single}. This approach allows one to use only a single measurement qubit and can lead to measurement protocols that require only a constant time to implement. However, the optimal control approach also suffers from two important problems both of which originate from the lack of a discernible structure in the optimal pulse shapes.
First, the optimized pulse shapes for different $r$ and $k$ do not have a clear relationship to each other (see for example Ref. \cite{curtis2021single} ). Hence, the optimization of the pulse shapes should be repeated for each $r$ and $k$ independently.
Second, finding optimal control pulses becomes increasingly more difficult for systems with a larger number of excitations because the optimization of the controls requires classical simulations of the measured system. 
Resolving these two problems would greatly improve 
the scalability and feasibility of generalized parity measurements, dramatically increasing their impact on near-term quantum technologies.

Here, we resolve the above problems 
using a toolset developed for the design of efficient quantum algorithms. In addition, we numerically demonstrate the effectiveness of our approach in preparing $>90\%$ fidelity multi-component cat states \cite{michael2016new,albert2016holonomic} of at up to $400$ photons on the current circuit quantum electrodynamics (cQED) platforms and discuss the main limitations of the implementation for larger photon numbers.

Our approach relies on a quantum algorithmic toolbox called Quantum Signal Processing (QSP) \cite{low2016methodology,low2019hamiltonian,gilyen2019quantum,lloyd2021hamiltonian}, which allows us to synthesize an appropriate unitary acting on both the measurement qubit $\mathcal{Q}$ and the measured system $\mathcal{S}$, independently of the dimension $D_{\mathcal{S}}$ of $\mathcal{S}$. In particular, we use QSP-based synthesis to construct a control protocol that approximately realizes the following unitary 
\begin{align}
\nonumber\hat{U}_{r,k} &= \mathbf{I}_{\mathcal{Q}} \otimes \sum_{\ket{m,j} \in S_{r,k}} \ket{m,j}\bra{m,j}\\
&+ \sum_{\ket{m,j} \notin S_{r,k}} iX_{ \eta_m } \otimes \ket{m,j}\bra{m_j},
\label{eq:QSPUni}
\end{align}
for \textit{all} $r$ and $k$. Here, $\mathbf{I}_{\mathcal{Q}}$ is the identity operator acting on $\mathcal{Q}$, the set $\{\ket{m,j}\}$  represents $n_m$ orthogonal states with $m$ excitations; and finally we define $X_{\eta_m}\equiv e^{i\eta_m/2 Z}Xe^{-i\eta_m/2 Z}$ with Pauli operators $X$ and $Z$ acting on $\mathcal{Q}$ and phase factors $e^{i\eta_m}$ that in general depend on the excitation number $m$. Intuitively, $\hat{U}_{r,k}$ flips the state of $\mathcal{Q}$ and imparts a phase shift only on the components $\ket{m,j}\notin S_{r,k}$. GP$_{r,k}$ can be implemented using 2 consecutive $\hat{U}_{r,k}$ each followed by a single-qubit measurement on the computational basis \cite{supplementary}. The main result of this work is an analytical expression describing a control protocol that i) results in a good approximation of $\hat{U}_{r,k}$ for all $r$ and $k$, ii) does not require numerical optimization for different $r$ and $k$, and iii) requires $\mathcal{Q}$ to interact with $\mathcal{S}$ only for a constant time that is independent of the size of $\mathcal{S}$. 

Our protocol can be used to measure both bosonic and qubit-based systems, as long as the interaction between $\mathcal{S}$ and $\mathcal{Q}$ is a QND Hamiltonian of the form $X\otimes H_S$, where the \textit{signal} Hamiltonian is defined as $H_S \equiv \chi \sum_{m}\sum_{j}^{n_m} m  \ket{m,j}\bra{m,j}$, with $\chi$ denoting the interaction strength between $\mathcal{S}$ and $\mathcal{Q}$.
As long as the system's evolution conserves the number of excitations, measuring $\mathcal{Q}$ induces a Quantum Non-Demolition (QND) measurement on $\mathcal{S}$ \cite{wiseman2009quantum, hann2018robust}, which can be repeated to improve the measurement fidelity.

\begin{figure}
    \centering
    \includegraphics[width=0.40\textwidth]{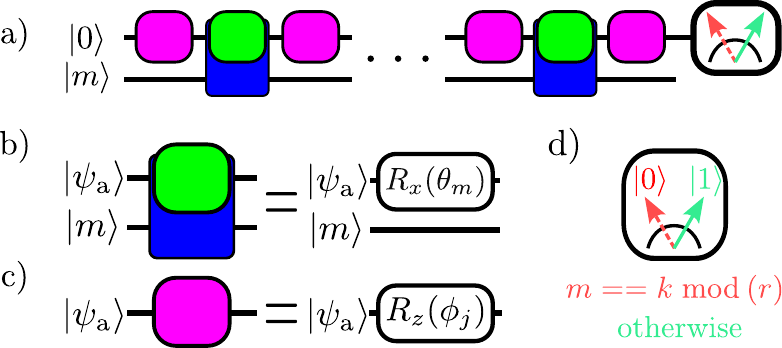}
    \caption{ a) Schematic representation of the proposed implementation of GP$_{r,k}$. The implementation requires only a single ancillary qubit $\mathcal{Q}$, which goes through repeated interactions with the measured system $\mathcal{S}$. b) The interaction Hamiltonian (in blue and green) amounts to rotating $\mathcal{Q}$ along a given axis by an angle that depends on the number of excitations in $\mathcal{S}$. c) The protocol also requires single-qubit processing unitaries (in purple) along a direction orthogonal to that induced by the interaction between $\mathcal{S}$ and $\mathcal{Q}$. d) The task of designing an implementation of the generalized parity measurement is then reduced to finding the processing angles $\{\phi_j\}$ such that at the end of the implementation, measuring $\mathcal{Q}$ in $\ket{0}$ and $\ket{1}$ projects $\mathcal{S}$ onto $S_{r,k}$ and its complement, respectively.}  
    \label{fig:QSP}
\end{figure}

Let us finally elaborate on our approach to synthesizing $\hat{U}_{r,k}$. Quantum Signal Processing (QSP) has become a general framework for designing optimal implementations of quantum algorithms \cite{low2016methodology,low2019hamiltonian,gilyen2019quantum,lloyd2021hamiltonian}. Despite its expressiveness, the core of QSP (and the way we use it approximate to $\hat{U}_{r,k}$) can be understood by considering a composite pulse sequence acting on a single qubit \cite{low2016methodology,low2017optimal,low2017optimal}. 
To this end, consider the following composite pulse sequence of depth $d$ acting on a single qubit \cite{low2014optimal,low2016methodology}
\begin{align}
\nonumber U_{\vec{\Phi}}(\theta) &= e^{i \Phi_0 Z}e^{i\theta X}e^{i \Phi_1 Z}e^{i\theta X}\cdots e^{i\theta X}e^{i \Phi_d Z}
\label{eq:QSP}
\end{align}
where $\theta \in \mathbf{R}$, and $\Phi_j\equiv \vec{\Phi}_j\in \mathbf{R}$ is a so-called processing phase associated with the processing unitary $e^{i\Phi_j Z}$ \cite{low2017optimal}. Most importantly for this work, the fundamental theorem of QSP says that $U_{\vec{\Phi}}(\theta)$ has the representation
\begin{align}
U_{\vec{\Phi}}(\theta)&= \left(\begin{array}{cc} P_{\vec{\Phi}}(\cos{\theta})& i \sin{(\theta)}Q_{\vec{\Phi}}(\cos{\theta}) \\ i \sin{(\theta)}Q_{\vec{\Phi}}^*(\cos{\theta}) & P_{\vec{\Phi}}^*(\cos{\theta})\end{array}\right),
\end{align}
where the matrix elements are complex-valued fixed-parity trigonometric polynomials  
whose coefficients are determined by $\vec{\Phi}$ \cite{low2016methodology,supplementary}.

It is straightforward to establish the relationship between Eq. ($\ref{eq:QSP}$) and the desired unitary $\hat{U}_{r,k}$ in Eq. ($\ref{eq:QSPUni}$). To this end, consider the unitary  $\tilde{U}_{\vec{\Phi}}(H_S,r,k)$ which has the same form as Eq. ($\ref{eq:QSP}$) but with the following replacement
\begin{align}
e^{i\theta X}\leftarrow V_S(H_{S},r,k) \equiv \exp{\left[-i \frac{\pi}{r} X \otimes ( H_{S}/\chi - k) \right]}.
\label{eq:subs}
\end{align}
Fig. $\ref{fig:QSP}$ depicts the schematic circuit of the control protocol implementing the QSP unitary.
Then, given an eigenstate $\ket{m,j}$ of $H_S$ with an eigenvalue $
\chi m$, the following relation is easily verified
\begin{align}
\tilde{U}_{\vec{\Phi}}(H_S,r,k) (\mathbf{I} \otimes \ket{m,j})= U_{\vec{\Phi}}\left[\theta_{m}^{r,k}\right]
\otimes\ket{m,j}
\label{eq:QSP_Full}
\end{align}
where $\theta_{m}^{r,k} \equiv \frac{\pi}{r} (m - k) $.
Consequently, there exists a protocol $\tilde{U}_{\vec{\Phi}_r}(H_S,r,k)$ that approximates the desired unitary $\hat{U}_{r,k}$ if
\begin{align}
\bra{0}U_{\vec{\Phi}}(\theta_{m}^{r,k})\ket{0}
\approx G_r(\cos \theta_{m}^{r,k} ) \equiv \bigg\{ \begin{array}{cc}1& m - k  = 0 \mod r \\ 0& \rm{otherwise}
\end{array},
\label{eq:Target}
\end{align}
where we chose the target polynomial $G_r\in \mathbf{R}[x]$. In the Supplementary Material  \cite{supplementary}, we show that this reality condition is satisfied for a modified protocol $U^{\mathbf{R}}_{\vec{\Phi}_{\rm sym}} \equiv -iX{\rm H}U_{\vec{\Phi}_{\rm sym}}{\rm H}$, where the phases $\vec{\Phi}_{\rm sym}$ are restricted to be symmetric [i.e.,
$\vec{\Phi}_{\rm sym} = (\phi_0,\phi_1,\cdots,\phi_1,\phi_0)]$. In particular, we show that $\bra{0}U^{\mathbf{R}}_{\vec{\Phi}_{\rm sym}}\ket{0}=\mathrm{Re}\left(P_{\vec{\Phi}_{\rm sym}}\right)$.

\begin{figure}
    \centering \includegraphics[width=0.48\textwidth]{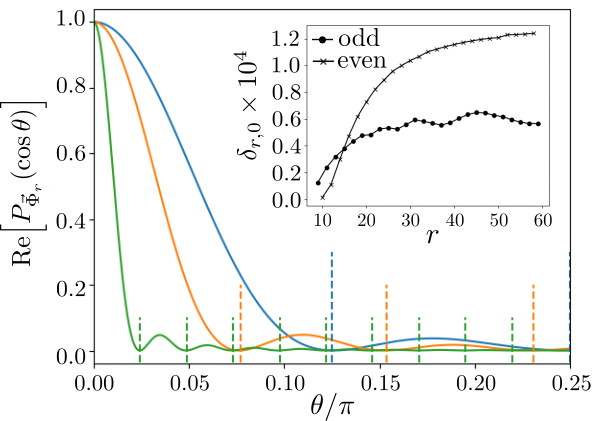}
    \caption{ The polynomial transformation ${\rm Re}\left[P_{\vec{\Phi}_r}\left(\cos \theta\right)\right]$ for $r = 8,13$ and $41$ using the approximate QSP angles in Eq. ($\ref{eq:AngleFormula}$). The vertical dashed lines in matching colors indicate the values of $\theta$ corresponding to $\theta_{m}^{r,k}$. The ideal polynomial $G_r(\cos{\theta_{m}^{r,k}})$ takes the value $0$ for $m  = k\,{\rm mod}\,r$ and  $1$ for $m \neq k\,{\rm mod}\,r$. Inset: the approximation error $\delta_{r,0}\times 10^4$ for $6>r>60$ (see main text).}
    \label{fig:Filter_Result}
\end{figure}

The important advantage of restricting ourselves to symmetric QSP protocols is that it allows us to leverage numerical optimization to determine the symmetric phase factors $\vec{\Phi}_{r}$ to approximate $G_r$ using numerical optimization described in Ref. \cite{wang2022energy,dong2022ground,supplementary}. Performing the numerical optimization for a few $r<30$ is sufficient for an analytical fit that works well for all $r$ considered in this work (see Fig. $\ref{fig:Filter_Result}$).
In particular, the phases $\vec{\Phi}_{r} = \left(\phi_{r}^{\rm edge}, \vec{\Phi}_{r}^{\rm bulk},\phi_{r}^{\rm edge}\right)$ for approximating $G_r$ are
\begin{align}
\nonumber 
\left(\vec{\Phi}_{r}^{\rm bulk}\right)_j &= \phi^{\rm bulk}_{r}(x_j) =
\frac{4}{\pi r} \left[1+ \frac{1}{2}\cos{\left(\frac{4}{r}x_j\right)} \right]\\
\phi_{r}^{\rm edge} &= \frac{1}{2}\left[\frac{\pi}{2} - \sum_{j=1}^{r-1}\left(\vec{\Phi}_{r}^{\rm bulk}\right)_j\right]
\label{eq:AngleFormula}
\end{align}
where $x_j = j - \ceil{\frac{r}{2}} $ with $j\in \left\{0,\cdots,2\times\left(\ceil{\frac{r}{2}}-1\right)\right\}$. 
The function ${\rm Re}\left[P_{\vec{\Phi}_r}(\cos{\theta})\right]$ for $r = 8,13,$ and $41$ are shown in Fig. $\ref{fig:Filter_Result}$. 
We evaluate the approximation error $\delta_{r,k}$ by comparing the ideal measurement operators $M_{\sigma_1}^{r,k}$ to the QSP-based measurement operators  $\bar{M}_{\sigma_2\sigma_1}^{r,k}\equiv \Pi_{\sigma_2}\tilde{U}^{\mathbf{R}}_{\vec{\Phi}_r}(H_S,r,k)\Pi_{\sigma_1}\tilde{U}^{\mathbf{R}}_{\vec{\Phi}_r}(H_S,r,k)$, conditioned on the measurement outcome $\sigma_1\sigma_2$ (here, $\tilde{U}^{\mathbf{R}}_{\vec{\Phi}_r}\equiv -iX{\rm H}U_{\vec{\Phi}_{\rm sym}}{\rm H}$). Formally, we calculate $$
\delta_{r,k} \equiv \frac{1}{D_{\mathcal{S}}}\left[{\rm Tr}\left(F_{0}^{r,k}\bar{F}_{00}^{r,k}\right) + {\rm Tr}\left(F_{1}^{r,k}\bar{F}_{10}^{r,k}\right)\right],$$
where 
$F_{\sigma}^{r,k}\equiv \left(M_{\sigma}^{rk}\right)^{\dagger}M_{\sigma}^{rk}$ and $\bar{F}_{\sigma_1\sigma_2}^{r,k}\equiv \left(\bar{M}_{\sigma_1\sigma_2}^{rk}\right)^{\dagger}\bar{M}_{\sigma_1\sigma_2}^{rk}$ are the relevant positive operator valued measure (POVM) operators.

Crucially, the sum $\sum_{j=1}^{r+1}\left(\vec{\Phi}_r\right)_j = \pi/2$ is a constant (as expected from the analytical bound provided in Ref. \cite{wang2022energy}).
As a result, given access to an evolution generated by $X\otimes H_S$, the total time required to implement the generalized parity measurement ${\rm GP}_{r,k}$ scales only as $O(k)$. Better still, when considering physical platforms for which the single-qubit Rabi frequency is much faster than $1/\chi$, the time to implement ${\rm GP}_{r,k}$ is practically constant because the $k$-dependent evolution in Eq. ($\ref{eq:subs}$) is a single-qubit rotation on the $\mathcal{Q}$. 

Before discussing the preparation of large cat states in a realistic setting, we note that because $\left|\phi_{r}^{\rm bulk}(x_j)\right|\leq 6/(\pi r)$ the susceptibility of the protocol to coherent errors (e.g., timing errors) increases linearly with $r$, thereby reducing the impact of our work for practical implementations with $r$ greater than a several hundred. However, this challenge can be overcome for certain large $r$ using a sequence of $s=O(\log(r))$ generalized parity measurements $\{{\rm GP}_{r_i,k_i}\}$ that leverage the Chinese Remainder Theorem \cite{graham1994concrete} as well as the QND nature of ${\rm GP}_{r,k}$. In particular, given that the desired modulus $r$ is a square free integer with $s$ non-repeating prime factors $\{r_i\}$, then ${\rm GP}_{r,k}$ can be implemented through a unique sequence of measurements $\prod_{j=1}^{s} {\rm GP}_{r_i,k_i}$ where $k_i = k \mod{r_i}$. Since $\max_i r_i$ and $\max_i k_i$ scale as $ O(r^{1/s})$, the susceptibility of the generalized measurement protocol to coherent errors increases only as $r^{1/s}$, at the cost of a total measurement time of $O(s r^{1/s})$. 

\textit{Preparing multi-component cat-states with cavity QED:} 
The protocol for preparing an $r$-component cat-state $\ket{\phi^{(l,\bar{n})}_{\rm des}}$ with an average photon number $\bar{n}$ consists of two steps. First, we prepare an initial state $\ket{\psi_{\rm init}}\equiv \ket{0}\ket{\alpha = \sqrt{\bar{n}}}$, where $\ket{0}$ is the initial state of $\mathcal{Q}$ and $\ket{\alpha = \sqrt{\bar{n}}}$ is the $\bar{n}$ photon coherent state of $\mathcal{S}$. Second, we apply $\tilde{U}^{\mathbf{R}}_{\vec{\Phi}_l}(H_S,l,0)$ on $\ket{\psi_{\rm init}}$ and measure $\mathcal{Q}$ on the computational basis. The state preparation \textit{succeeds} with probability $p_{\rm succ}$ if $\mathcal{Q}$ is projected onto $\ket{0}$. We emphasize that this protocol is also applicable to qubit states by replacing the photonic coherent state with a spin-coherent state.  

Neglecting error mechanisms, the probability $p^{0}_{\rm succ}$ of successful state preparation can be calculated from the overlap between $\ket{\psi_{\rm init}}$ and $\ket{\phi^{(l,\bar{n})}_{\rm des}}$. A simple numerical evaluation of $p^{0}_{\rm succ}$ yields $p^{0}_{\rm succ}\approx l^{-1}$ for $l\leq 2\sqrt{\bar{n}}$. In contrast, in the large-$l$ limit ($l > 2\sqrt{\bar{n}}$), $\bar{n}\,{\rm mod}\,l = 0$ implies that   $\ket{\phi^{(l,\bar{n})}_{\rm des}}\approx \ket{\bar{n}}$  where $\ket{\bar{n}}$ is a photon-number state with $\bar{n}$ photons. Under these conditions, we find $p^{0}_{\rm succ}\approx|\bra{\bar{n}kabalsky1989
}\alpha=\sqrt{\bar{n}}\rangle|^2\approx (2.5 \sqrt{\bar{n}})^{-1}$  \cite{supplementary}. The behavior of $p^{0}_{\rm succ}$ together with the fact that our state preparation protocol requires a constant time entails that i) the time complexity for successfully preparing $l$-component cat states is $O(l)$ for $l\leq \sqrt{\bar{n}}$ (neglecting the preparation of the initial coherent state) and ii) the time complexity for preparing photon states is $O(\sqrt{\bar{n}})$. Remarkably, if we use our protocol to prepare of Dicke states with $\bar{n}$ excitations in qubit systems, the latter result implies that the time complexity of our \textit{nondeterministic} scheme matches that conjectured to be optimal for the \textit{deterministic} preparation of Dicke states using a quantum processor with all-to-all coupling \cite{bartschi2022short,krastanov2015universal}.

\begin{figure}
    \centering \includegraphics[trim = 0mm 0mm 0mm 0mm, clip, width=0.5\textwidth]{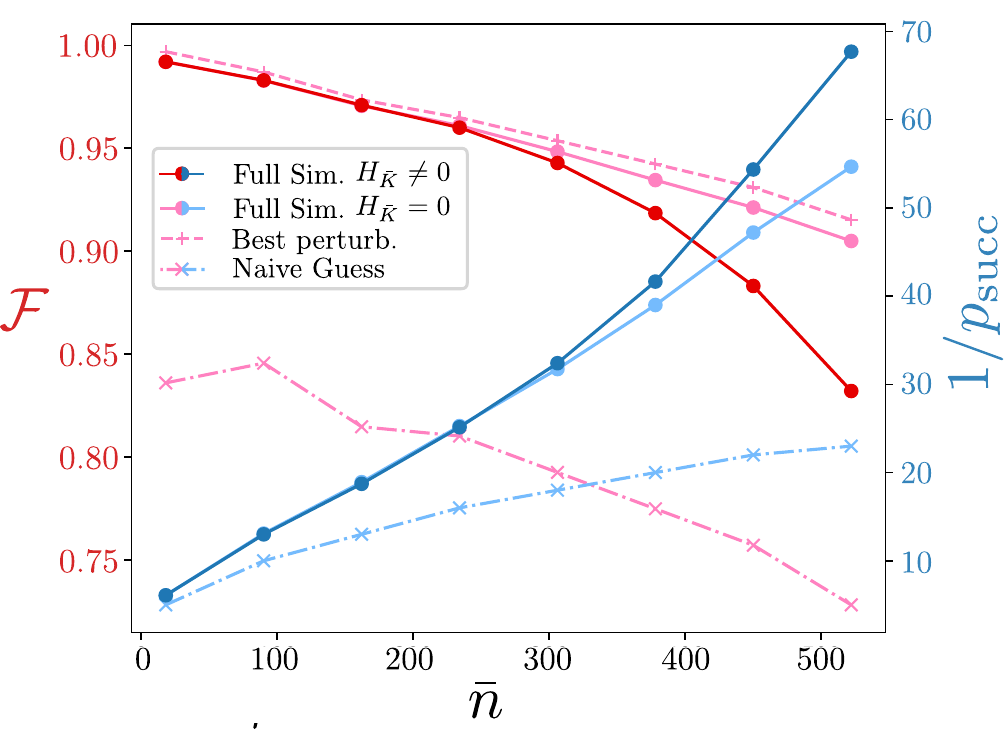}
    \caption{The fidelity $\mathcal{F}$ and the inverse success probability $1/p_{\rm succ}$, for a successful 3-times repeated $\bar{n}$ photon $\ceil{\sqrt{\bar{n}}}$-component cat state preparation protocol,. The full lines correspond to full master equation simulations, with red and pink (blue and light blue) full disks representing the data resulting from simulations with $H_{\bar{K}}\neq 0$ and $H_{\bar{K}}= 0$, respectively.  
    By repeating the state-preparation protocol 3 times, we obtain a fidelity that closely approximates the theoretical approximation given by considering only cavity decay (pink dashed line) for $\bar{n}\leq 300$, substantially improving the naive guess for the fidelity of a 1-time repeated state preparation protocol \cite{supplementary} (pink dashed-dotted line).  
    The maximal inverse success probability $1/p^0_{\rm succ}$ for preparing a $\ceil{\sqrt{\bar{n}}}$-component $\bar{n}$-photon cat state (blue dotted-dashed line) 
    scales with the square root of the number of legs (see main text). Repeating the state preparation 3 times reduces $p_{\rm succ}^0$ for $\bar{n}=378$ by $\approx 1/2$.}
    \label{fig:FidResults}
\end{figure}

Next, we turn our attention to a concrete implementation of the cat state preparation protocol on a superconducting cavity QED device. In order to implement the unitary $\tilde{U}^{\mathbf{R}}_{\vec{\Phi}_l
}$ in Eq. ($\ref{eq:QSP_Full}$), we can use the native qubit-cavity coupling Hamiltonian $H_{\chi} \equiv \chi Z\otimes n_C$ in the highly dispersive regime (here $n_C$ denotes the cavity-photon number operator)\cite{boissonneault2009dispersive,Sivak2022}. However, when the number of photons is large, realizing the desired dynamics becomes difficult due to multiple sources of unwanted coherent processes.
A crucial obstacle against implementing the unitary $\tilde{U}_{\vec{\Phi}_l
}$ is the requirement to cancel the qubit-cavity interactions during the implementation of the QSP processing unitaries. While 
$H_{\chi}$ can be switched off using additional drives \cite{rosenblum2018fault},  
our numerical results indicate that the adverse effects of always-on $H_{\chi}$ can be drastically reduced through the following replacement of the single-qubit processing unitary
\begin{align}
e^{i\phi_j X} \leftarrow e^{i(\phi_jX-\bar{n}\chi t_{\phi_j}Z)},
\label{eq:Cancellation}
\end{align}
where $t_{\phi_j}$ is the time required to implement $e^{i\phi_j X}$. In our simulations, we observe that this replacement does not result in a significant error for non-classical photonic states with at least up to 500 photons (see the pink full line in Fig. $\ref{fig:FidResults}$). The effectiveness of the replacement in Eq. ($\ref{eq:Cancellation}$) can be understood by noting that the measurement protocol is QND and the system is initialized in a coherent state.  

Another source of coherent errors is the presence of the always-on qubit-state-dependent Kerr nonlinearity  $H_{\bar{K}}\propto \bar{K}\sigma_z n_C^2$, where $\bar{K}\ll \chi$ \cite{blais2021circuit,zhang2022drive,supplementary}. The effects of this nonlinear coupling can also be reduced by modifying the single-qubit rotation in the same way as in Eq. ($\ref{eq:Cancellation}$). However, $H_{\bar{K}}$ also introduces coherent errors in the implementation of $V_{\rm S}(H_{\rm S},l,k)$ in Eq. ($\ref{eq:subs}$). These errors are the main limitation of the QSP-based cat-state preparation protocol. We find that, for the parameters we chose \cite{supplementary}, the effect of the always-on $H_{\bar{K}}$ becomes relevant for $\bar{n} > 350$ (see Fig. $\ref{fig:FidResults}$).  Finally, we emphasize that the effective Kerr nonlinearity $H_K\propto n_C^2$ commutes with all components of the QSP-based generalized parity measurement. Therefore, the evolution due to the Kerr nonlinearity can be cancelled before or after implementing GP$_{r,k}$. This cancellation could be done either in post-processing or by changing the sign of the cavity Kerr non-linearity by driving the qubit close to resonance \cite{zhang2022drive,supplementary}. Hence, in the following, we report fidelities that do not include the effect of $H_K$.

In Fig. $\ref{fig:FidResults}$, we present the inverse success probability $p_{\rm succ}^{-1}$ and the fidelity $\mathcal{F} \equiv \sqrt{\bra{\phi_{\rm des}}\sigma_{\rm final}\ket{\phi_{\rm des}}}$ of the post-successful-measurement state $\sigma_{\rm final}$, 
in the presence of all relevant decoherence channels in the superconducting cavity QED setting. For simplicity, we consider only the preparation of $\ceil{\sqrt{\bar{n}}}$-component cat states with average photon number $\bar{n}$.
The most important system parameters for understanding our results are the cavity lifetime $T_c$ and the qubit-cavity coupling parameter $\chi$.
For the results presented here, we chose $T_c = 25$ ms \cite{milul2023superconducting} and $\chi =2\pi \times 41$ kHz \cite{Sivak2022} (see \cite{supplementary} for all system parameters).

Surprisingly, the QSP-based state preparation protocol further reduces the photon-loss-induced infidelity of the post-measurement state by $\approx 1/2$. We develop a perturbative expansion of the stochastic wavefunction trajectories \cite{dalibard1992wave,castin2008wave} to analyze the mechanism behind the robustness of our scheme to the main decoherence channels in the Supplementary Materials \cite{supplementary}. We also find that repeating the state preparation protocol 3 times brings the fidelity of the resulting cat state very close to that predicted by the first-order perturbation theory that \textit{only} takes into account the cavity decay processes \cite{supplementary}. We find that the proposed state preparation scheme can be used to prepare 20-component cat states with $\bar{n}=378$ photons with $\mathcal{F}\approx 92\%$ and $p_{\rm succ}= 2.4\%$. 
Finally, our numerical results also show that the fidelity of the state preparation does not change substantially when the number of cat-state components is increased, even in the presence of coherent and incoherent errors. 
In particular, we show that photon number state with $\bar{n}=376$ photons can be prepared with $\approx 94\%$ fidelity and $\approx 1.1\%$ success probability by applying our protocol with $r=94$ to an initial coherent state with $\bar{n}=378$ photons \cite{supplementary}.

In this letter, we have presented a way of leveraging the optimal resource usage of the algorithmic framework of QSP to i) efficiently implement generalized parity measurements and ii) non-deterministically prepare large multi-component cat states. Our scheme can be translated to various other experimental platforms that allow access to constant time implementations of $e^{-i t X \otimes H_S}$, such as neutral atoms \cite{molmer2011efficient,isenhower2011multibit,young2020asymmetric}, circuit acustodynamics \cite{von2022parity}, and other cavity QED settings \cite{zhu2016measurement,anikeeva2021number}  
We have shown that the efficiency of our protocol results in high-fidelity preparation of non-classical states with a large number of photons in the superconducting cavity QED setting. Further developments in reducing the cavity decay rate \cite{romanenko2020three} and canceling unwanted coherent terms can greatly improve this result. Our simulations for $H_{\bar{K}}=0$ shows that cat states of up to $\bar{n}=500$ photons can be prepared with fidelities limited only by the cavity decay rate \cite{supplementary}. 
For future work, it will be important to consider replacing the signal Hamiltonian with other experimentally realized operators (such as the cavity quadrature operators) and address the question whether 
optimizing the processing angles of the QSP protocol can reduce the adverse effects of system-specific decoherence channels.

\begin{acknowledgments}
 \textit{Acknowledgments:} The author is deeply grateful for enlightening discussions with Marek Pechal on the experimental implementation of generalized parity measurements in cavity QED and Isaac Chuang, Arkopal Dutt, and Sho Sugiura, on Quantum Signal Processing during the initial stages of this project. The author acknowledges the financial and technical support of NTT PHI Labs.
\end{acknowledgments}


\end{document}


\preprint{APS/123-QED}

\title{Supplementary Material for: Generalized parity measurements and efficient large multi-component cat state Preparation with Quantum Signal Processing}

\author{Sina Zeytino\u glu }
\affiliation{
Physics and Informatics Laboratory, NTT Research, Inc.,940 Stewart Dr., Sunnyvale, California, 94085, USA }

\maketitle

\section{Introduction to symmetric QSP}

Here, we state the formal results describing the framework of symmetric Quantum Signal Processing \cite{low2017optimal,wang2021preparing,wang2022energy} for a single qubit. The structure of single qubit symmetric QSP is 
\begin{align}
 U_{\vec{\Phi}}(\theta) &= e^{i \Phi_0 Z}e^{i\theta X}e^{i \Phi_1 Z}e^{i\theta X}\cdots e^{i\theta X}e^{i \Phi_d Z},
 \label{eq:single_qubit_QSP_supp}
\end{align}
where the phases $\Phi_j$ are symmetric in the sense that 
\begin{align}
\vec{\Phi} = (\phi_0,\phi_1,\cdots,\phi_1,\phi_0) \in [-\pi,\pi)^{d+1}.
\end{align}
In this work, we follow \cite{wang2022energy} and take the domain of the symmetric phase factors to be 
\begin{align}
D_d =\bigg\{  
\begin{array}{cc} [-\frac{\pi}{2},\frac{\pi}{2})^{d/2}\times[-\pi,\pi)\times [-\frac{\pi}{2},\frac{\pi}{2})^{d/2} & d\,\, {\rm is \,\, even} \\
{[}-\pi/2 , \pi/2) ^{d+1} & d \,\, \mathrm{is \,\, odd} \end{array}
\end{align}

The core theorem of symmetric QSP is the following:
\begin{theorem} [Theorem 1, \cite{wang2022energy}] Consider any polynomial $P\in \mathbf{C}[x]$ and $Q\in\mathbf{R}[x]$, satisfying the following conditions
\begin{enumerate}
\item deg$(P)=d$ and deg$(Q)=d-1$ 
\item par$(P)=d \mod 2$ and par$(P)=(d-1) \mod 2$ 
\item $\forall x \in [-1,1]: |P(x)|^2+(1-x^2)|Q(x)|^2=1$
\item if $d\in$ odd, then the leading coefficient of $Q$ is positive
\end{enumerate}
There exists a unique set of symmetric phase factors $\vec{\Phi}_{(P,Q)}\equiv (\phi_0,\phi_1,\cdots,\phi_1,\phi_0)\in D_d$ such that 
\begin{align}
 U_{\vec{\Phi}_{(P,Q)}}(\theta) =\left( \begin{array}{cc} P(x) & iQ(x)\sqrt{1-x^2}\\ iQ(x)\sqrt{1-x^2} & P^*(x) \end{array}\right)= \mathrm{Re}[P(x)] \mathbf{I} + i \mathrm{Im}[P(x)] Z + i Q(x)\sqrt{1-x^2} X,
\end{align}
where $x\equiv \cos{\theta}$.
\end{theorem}
In the above theorem statement $\mathrm{Re}(y)$ and $\mathrm{Im}(y)$ denote the real and imaginary part of $y\in \mathbf{C}$, and ${\rm deg}$ and ${\rm par}$ denote the degree and parity of polynomial functions, respectively.

To design the generalized parity measurement ${\rm GP}_{r,k}$, we would like to realize a modified unitary  $\bar{U}_{\vec{\Phi}_{(P,Q)}}(\theta)$ that obeys  $$\bra{0}\bar{U}_{\vec{\Phi}_{(P,Q)}}(\theta)\ket{0} \in \mathbf{R}[x].$$ To satisfy this constraint, we leverage the fact that our discussion in the main text does not require the off diagonal matrix elements to be purely imaginary. In particular, we simply observe that, given any $\vec{\Phi}$ and the associated polynomials $P$ and $Q$ satisfying the conditions of the Theorem 1,
\begin{align}
i X He^{-i\pi/4 Z} U_{\vec{\Phi}_{(P,Q)}}(\theta)e^{-i\pi/4 Z}H = \left[{\rm Re}(P(x))\mathbf{I}+i({\rm Im}(P(x))X - Q(x)\sqrt{1-x^2}Y)\right].
\end{align}
In other words, $i\bra{0}X {\rm H}e^{-i\pi/4 Z} U_{\vec{\Phi}_{(P,Q)}}(\theta)e^{-i\pi/4 Z} {\rm H}\ket{0} \in \mathbf{R}[x]$. Notice that the single-qubit $Z$ rotations before and after $U_{\vec{\Phi}_{(P,Q)} }$ can be incorporated into the QSP phase sequence to obtain a new symmetric phase sequence $\vec{\bar{\Phi}}_{(P,Q)} \equiv (\phi_0-\pi/4,\phi_1,\cdots,\phi_1,\phi_{0}-\pi/4)$. Thus in the following, we will be interested in unitaries  $U^{(\mathbf{R})}_{\vec{\Phi}_{\rm sym}}\equiv iX{\rm H}U_{\vec{\Phi}_{\rm sym}}{\rm H}X$, where $\vec{\Phi}_{\rm sym}$ is symmetric.  

\textit{Finding the symmetric phase sequences to implement $\hat{U}_{r,k}$:} To obtain the analytical results for symmetric phase factors $\vec{\Phi}_{r}$ associated with target functions $G_r(x)\in \mathbf{R}[x]$, such that 
\begin{align}
\bra{0}U^{(\mathbf{R})}_{\vec{\Phi}_r}\ket{0}\approx G_r(\cos \theta_{m}^{r,k} ) \equiv \bigg\{ \begin{array}{cc}1& m - k  = 0 \mod r \\ 0& \rm{otherwise}
\end{array},
\label{eq:Target}
\end{align}
(where $\theta_{m}^{r,k}\equiv \frac{\pi}{r}(m-k)$ as in the main text), we use the numerical optimization approach first introduced by of Ref. \cite{wang2021preparing,wang2022energy} for symmetric phase sequences. Specifically, we numerically calculate the phase factors for polynomial approximations of $G_r$ up to $r<30$ and $r\in even$ by minimizing the following cost function 
\begin{align}
F_{\vec{\Phi}} = \frac{1}{\tilde{d}}\sum_{j}^{\tilde{d}} \left| {\rm Re}(P_{\vec{\Phi} }(x_j))-G_r(x) \right|^2,
\end{align}
where $\tilde{d} \equiv \ceil{\frac{d+1}{2}}$, and $x_j \equiv \cos{\left(\frac{2k-1}{4\tilde{d}}\right)}$ are the nodes of an order-$d$ Chebyshev polynomial of the first kind.  By fitting the ordered elements of the optimal phase sequences $\vec{\Phi}_r$ to trigonometric functions and extending to $r\in odd$, we find the simple expressions presented in the main text. That is, $\vec{\Phi}_r = (\phi_r^{\rm edge},\vec{\Phi}_r^{\rm bulk}, \phi_r^{\rm edge})$, with 
\begin{align}
\nonumber 
\left(\vec{\Phi}_{r}^{\rm bulk}\right)_j &= \phi^{\rm bulk}_{r}(x_j) =
\frac{4}{\pi r} \left[1+ \frac{1}{2}\cos{\left(\frac{4}{r}x_j\right)} \right]\\
\phi_{r}^{\rm edge} &= \frac{1}{2}\left[\frac{\pi}{2} - \sum_{j=1}^{r-1}\left(\vec{\Phi}_{r}^{\rm bulk}\right)_j\right]
\label{eq:AngleFormula}
\end{align}
where $x_j = j - \ceil{\frac{r}{2}} $ with $j\in \left\{0,\cdots,2\times\left(\ceil{\frac{r}{2}}-1\right)\right\}$.

\begin{figure}[h!]
    \centering
    \includegraphics[width=0.45\textwidth]{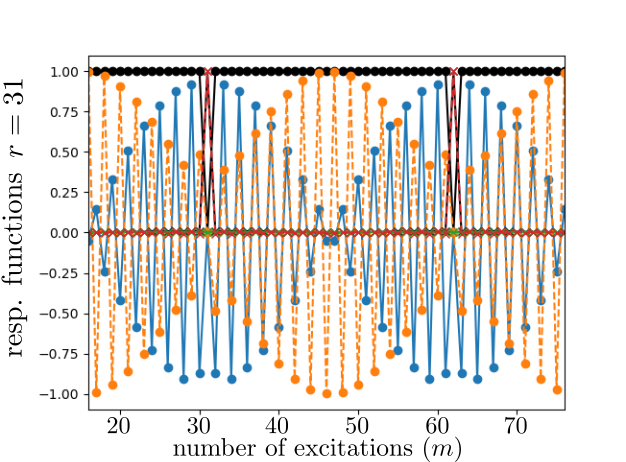}
    \includegraphics[width=0.45\textwidth]{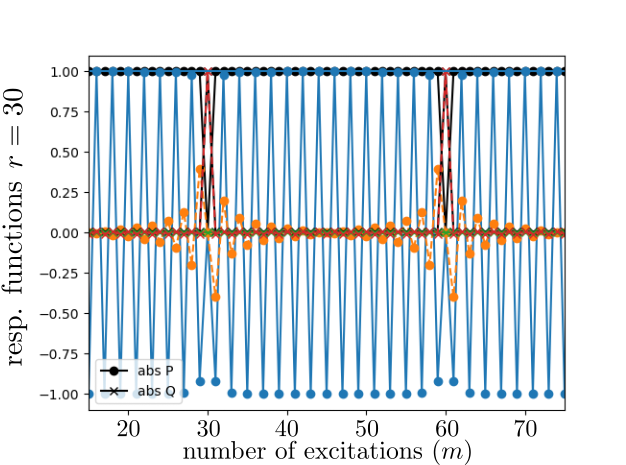}
    \caption{The real and imaginary parts of the response functions ${\rm Re}\left[R_{m,r}^{(10)}\right]$ (in blue disks), ${\rm Im}\left[R_{m,r}^{(10)}\right]$ (in yellow disks), and $R_{m,r}^{(00)}$ (red crosses) for $r=31$ and $r=30$ as a function of the number of excitations $m$ in the measured system. Observe that the response $R_{m,r}^{(00)}$ takes a real value close to 1 at $m= 0\,{\rm mod}\,r$, while $R_{m,r}^{(10)}$
    is $\approx 0$. On the other hand, when $m\neq 0\,{\rm mod}\,r$, $R_{m,r}^{(00)}$ is approximately 0 and $|R_{m,r}^{(10)}|$ takes a complex value whose absolute value is approximately 1 (black disks). Intriguingly, for $r=31$,  the phase winding of the response function $R_{m,r}^{(10)}$ is almost constant, except near $m= 0\,{\rm mod}\,r$.}
    \label{fig:odd_r}
\end{figure}

It is useful to provide further details regarding the phase sequences $\vec{\Phi}_r$ in Eq. ($\ref{eq:AngleFormula}$). The matrix elements (hereby called response functions) $R_{m,r}^{(00)}\equiv\bra{0}U^{\mathbf{R}}_{\vec{\Phi}_r}(\theta_{m}^{r,0})\ket{0}$ and 
$R_{m,r}^{(10)}\equiv\bra{1}U^{\mathbf{R}}_{\vec{\Phi}_r}(\theta_{m}^{r,0})\ket{0}$ as a function of $m$ are depicted in Fig. $\ref{fig:odd_r}$, for $r=31$ and $r=30$, respectively. To understand the behaviour of response functions, it is important to notice that independently of the parity of $r$, the number of processing phases is an odd number, thereby requiring the application of an even number of signal operators. Then, consider the case that the measured system has $m_0 = k \,{\rm mod}\,r$ number of excitations. In this case, $\theta_{m_0}^{r,k} = n\times\pi$, and $e^{i\theta_{m_0}^{r,k} X}=\pm \mathbf{I}$. Taking into account that the sum of the processing phases $\vec{\Phi}_r$ always add up to $\pi/2$, we obtain
\begin{align}
U^{\mathbf{R}}_{\vec{\Phi}_r}(\theta = \{0,\pi\}) = \mathbf{I},
\end{align}
Therefore, when the measured system has $m=k\mod r$ excitations, the measurement qubit initiated in the $\ket{0}$ state will transition to $\ket{0}$ without any overall phase shift at the end of the protocol, independently of the parity of $r$. On the other hand, when the measured system has $m \,\neq \,k \,{\rm mod}\, r$ excitations, we find the measurement qubit initiated in the $\ket{0}$ transition to $\ket{1}$ state with an additional phase shift that depends on $m$.  

\textit{Constructing GP$_{r,k}$:}
Next, we detail how to effectively remove the $m$-dependent phase in the off-diagonal elements of $U^{\mathbf{R}}_{\vec{\Phi}_{r}
}$ to realize the measurement operators $M_{0(1)}^{r,k}$ of the generalized parity measurement GP$_{r,k}$ 
\begin{align}
M_{0}^{r,k} = \ket{0}\bra{0}\otimes \Pi_{S_{r,k}} \quad \quad M_{1}^{r,k} = \ket{0}\bra{0}\otimes \Pi_{\notin S_{r,k}},
\end{align}
where $\Pi_{S_{r,k}}$ and $\Pi_{\notin S_{r,k}}$ are orthogonal projectors projecting onto subspaces $S_{r,k}$ and its compliment, respectively \footnote{We remind the reader that $S_{r,k}$ is the subspace of the measured system $\mathcal{S}$, which is spanned by all states with $m = k \mod r$ excitations.}. It is straightforward to show how a perfect generalized parity measurement ${\rm GP}_{r,k}$ can be implemented using only $\hat{U}_{r,k}$ and single-qubit measurements without the need for resetting or feedback, in a sense similar to the restless methods used in the calibration of circuit QED devices \cite{rol2017restless,werninghaus2021high}. To this end, consider the scenario where the measurement qubit $\mathcal{Q}$ is initialized in $\ket{0}$, and the measured system $\mathcal{S}$ is initiated in an arbitrary state. Evolving the combined system $\mathcal{Q} + \mathcal{S}$ with $\hat{U}_{r,k}$ and measuring the ancillary qubit in the computation basis is described by the measurement operators $\tilde{M}^{r,k}_{0(1)} \equiv \Pi_{0(1)}\hat{U}_{r,k}\Pi_0$, and $\tilde{M}_1 \equiv\Pi_1\hat{U}_{r,k}\Pi_0$, where $\Pi_{0(1)} = \ket{0(1)}\bra{0(1)}\otimes \mathbf{I}_{\mathcal{S}}$ project $\mathcal{Q}$ onto the $\ket{0}$ ($\ket{1}$) state. These measurement operators have the same form as $M_{0(1)}^{r,k}$, except the arbitrary phase factors $\{e^{i\eta_m}\}$. Yet the effect of $\{e^{i\eta_m}\}$ can be readily eliminated by simply repeating the measurement protocol twice without any need to reset the ancilla or apply feedback on the measurement results. In the absence of errors, only the $00$ and $10$  measurement outcomes are possible (each outcome precedes the one to its right). For the latter, the phase factors imparted in the first round are eliminated in the second and the post-measurement state is identical to that under the perfect generalized parity measurement operators $M^{r,k}_{0(1)}$.

\section{The  details for the simulation of the cavity QED system}

The superconducting cavity-QED platform that we consider is as described extensively in the literature (see Ref. \cite{blais2021circuit} and references therein). Here, we only provide a brief introduction. The cavity-QED Hamiltonian is most naturally studied using a microscopic Hamiltonian which describes two coupled driven oscillators. 
\begin{align}
H_{0} = \hbar \omega_q a^{\dagger}a + \hbar\omega_c b^{\dagger}b - \frac{E_C}{2} \left(a^{\dagger}a^{\dagger}aa\right) - \hbar g(a-a^{\dagger})(b-b^{\dagger}),
\end{align}
where $a$ and $b$ are the annihilation operators for a non-linear oscillator of frequency $\omega_q$ and a linear oscillator of frequency $\omega_c$, respectively. The coupling between the two oscillators is given by $\hbar g$, and $E_C$ represents the strength of the Kerr non-linearity for the non-linear cavity. Assuming $g\ll \omega_q, \omega_r$, the rotating wave approximation is valid, resulting in
\begin{align}
H_{\rm RWA} = \hbar \omega_q a^{\dagger}a + \hbar\omega_c b^{\dagger}b - \frac{E_C}{2} \left(a^{\dagger}a^{\dagger}aa\right) + \hbar g( ab^{\dagger} + a^{\dagger}b).
\end{align}
Finally, when the nonlinearity $E_C$ is large compared to $g$, it is reasonable to include only the lowest two states $\ket{e}$ and $\ket{g}$, corresponding to oscillator states with 1 and 0 excitations, respectively \cite{blais2020quantum}. This approximation results in a Jaynes-Cummings Hamiltonian
\begin{align}
H_{\rm JC}/\hbar = \frac{\omega_q}{2} \sigma_z + \omega_c b^{\dagger}b + g( \sigma_-b^{\dagger} + \sigma_+ b),
\label{eq:JCHam}
\end{align}
where $\sigma_z \equiv \ket{e}\bra{e} - \ket{g}\bra{g}$ and $\sigma_- \equiv \sigma_{+}^{\dagger}= \ket{g}\bra{e}$. From here on, we refer to the two-level subsystem of the non-linear oscillator as the qubit and the harmonic oscillator as the cavity. The coherent cavity and qubit drives are described as 
\begin{align}
 H_{\rm JC, d}(t)/\hbar &= \frac{\omega_q}{2} \sigma_z + \omega_c b^{\dagger}b + g( \sigma_-b^{\dagger} + \sigma_+ b)+ i\Omega_c(t) (b - b^{\dagger}) + \Omega_q(t) \sigma_x,
\end{align}
where $\sigma^x = \sigma^+ +\sigma^-$ and $\Omega_c(t)$ and $\Omega_q(t)$ are the complex-valued coherent drive amplitudes, respectively. The Jaynes-Cummings Hamiltonian is a valid description of the cavity-QED system as long as $|\Omega_{q}|,|\Omega_r|\ll E_C$. 

In order to obtain results that are relevant for real experiments, we should also include the decoherence effects due to the coupling of the qubit-cavity system to its environment. The most widely used model captures Markovian decoherence processes through a Lindblad equation of motion of the density matrix
\begin{align}
\partial_t \rho = -\frac{i}{\hbar}\left[H_{\rm JC, d},\rho \right] + \sum_{i=1}^{n_{\rm diss}}\mathcal{D}_{L_i}(\rho),
\end{align}
where we suppressed the time dependence for simplicity, and used $L_i$ to denote the jump operators associated with $n_{\rm diss}$ decoherence channels \cite{gardiner2004quantum}, and 
\begin{align}
\mathcal{D}_{L_i}(\rho) \equiv L_i \rho L_i^{\dagger} - \frac{1}{2}\{\rho,L_{i}^{\dagger}L_i\}  .
\end{align}
For the cavity-QED setup, the main decoherence channels are described through the following jump operators
\begin{align}
L_{c,r} \equiv \sqrt{\gamma_q}b \quad\quad
L_{q,r} \equiv \sqrt{\gamma_c}\sigma_- \quad\quad
L_{q,\phi} \equiv \sqrt{\frac{\gamma_{q,\phi}}{2}}\sigma_z,
\end{align}
where $L_{q,\phi}$ describes the qubit dephasing, while $L_{q,r}$ and $L_{c,r}$ describe the qubit and cavity relaxation, respectively. The bare qubit decoherence rates $\gamma_q$ and $\gamma_{q,\phi}$ are experimentally accessible in the absence of the cavity through the longitudinal and transverse lifetimes $T_1$ and $T^{\rm R}_2$, respectively
\begin{align}
T_1 & \approx \frac{1}{\gamma_q}\quad\quad
T^{\rm R}_2 \approx \frac{1}{\gamma_q/2+\gamma_{q,\phi}}.
\end{align}
On the other hand, the cavity longitudinal relaxation rate $\gamma_c$ can be experimentally measured by initializing the cavity in a coherent state or a single-photon Fock state, and measuring the number of photons in the cavity the as a function of time. The cavity relaxation models 
\begin{align}
n_C(t) = n_C(0) e^{-\gamma_c t},
\end{align}
where $n_C(t)$ is the number of cavity photons at time $t$, with $n_C(0) = b^{\dagger}b$.

\subsection{Dynamics in the dispersive regime}

Importantly, the Jaynes-Cummings type native interaction between the qubit and the resonator is not a Quantum Non-Demolition (QND) interaction with respect to the rest of the terms in $H_{\rm JC, d}$. However, in the parameter regime where $g \ll |\omega_r-\omega_q|\equiv \Delta$, the effective dynamics of the qubit-resonator system can be approximated by considering an the following effective Hamiltonian, where the interaction between the qubit and the cavity approximates a QND interaction \cite{blais2021circuit}.
\begin{align}
H_{\rm QND} = \tilde{\omega}_cb^{\dagger}b + \frac{\tilde{\omega}_q}{2}\sigma_z + \chi \sigma_z \otimes b^{\dagger}b + \frac{K_C}{2} (b^{\dagger}b )^2 (1-\eta \sigma_z) + {\rm higher\, order\, in\,}b^{\dagger}b
\label{eq:FullHamiltonian}
\end{align}
where 
\begin{align}
\tilde{\omega}_c &= \omega_c - \frac{g^2}{\Delta- E_C/\hbar}\quad\quad \tilde{\omega}_q = \omega_q + \frac{g^2}{\Delta} \quad\quad 
\chi = -\frac{g^2 E_C/\hbar}{\Delta (\Delta - E_C/\hbar)}\quad\quad K_C \approx -E_C\left(\frac{g}{\Delta} \right)^4 \quad \quad \eta = \frac{1}{2}\frac{9 E_C}{\Delta}.
\end{align}
In Eq. ($\ref{eq:FullHamiltonian}$), the terms higher order in the photon number operator $n_C\equiv b^{\dagger}b$ can be neglected as long as the number of photons is smaller than the critical photon number $n^e_{\rm crit}$. In the regime $E_C \ll \Delta$, Ref. \cite{eickbusch2022fast} showed that 
\begin{align}
n_{\rm crit}^e \approx \frac{1}{2}\frac{E_C}{6\chi} \approx \frac{1}{6}\frac{\Delta^2}{2g^2}.
\end{align} 

The transformation that results in the effective Hamiltonian $H_{\rm QND}$, also transforms the Lindbladian jump operators that describe the decoherence channels in terms of the transformed qubit and cavity degrees of freedom. In particular, the decoherence mechanisms in the dispersive regime are well captured by the following jump operators in the transformed frame \cite{boissonneault2009dispersive}
\begin{align}
\tilde{L}_{c,r} &\equiv \sqrt{\gamma_c + \gamma_{cq}}b 
\quad\quad
\tilde{L}_{q,r} \equiv \sqrt{\gamma_{q} + \gamma_{qc} }\sigma_- 
\quad\quad
\tilde{L}_{q,\phi} \equiv \sqrt{\frac{\gamma_{q,\phi}}{2}}\sigma_z \quad\quad
\tilde{L}_{c,\Delta,1} \equiv \sqrt{\gamma_{c,\Delta}}\sigma_-b^{\dagger}\quad\quad
\tilde{L}_{c,\Delta,2} \equiv \sqrt{\gamma_{c,\Delta}}\sigma_+ b,
\label{eq:qubitDeph}
\end{align}
where 
\begin{align}
\gamma_{cq} = \left(\frac{g}{\Delta}\right)^2\gamma_q \quad\quad \gamma_{qc} = \left(\frac{g}{\Delta}\right)^2\gamma_c\quad\quad
\gamma_{c,\Delta} = 2\left(\frac{g}{\Delta}\right)^2\gamma_{q,\phi} = \frac{1}{6 n_{\rm crit}^e}\gamma_{q,\phi}.
\end{align}
Intuitively, $\gamma_{qc}$ and $\gamma_{cq}$ denote the relaxation rates of the dressed qubit and cavity degrees of freedom in the transformed basis, while $\gamma_{c,\Delta}$ is the dephasing of the cavity photons due to their coupling to the qubit degree of freedom. In our numerical simulations, we only simulate the effective dynamics of the qubit-cavity system in the dispersive regime. The values of the effective decay rates $\gamma_c + \gamma_{cq}$ and $\gamma_q + \gamma_{qc}$ are directly accessible experimentally, in a setup where the qubit is coupled to the cavity.

We also emphasize that the jump operators that originate from the dephasing rate of the qubit results in decoherence processes that no longer respect the QND nature of the proposed generalized parity measurement, as these jump operators do not conserve the number of cavity photons. 

We summarize the  values of the model parameters that are consistent with the literature \cite{Sivak2022,milul2023superconducting,werninghaus2021leakage} in Table I. We emphasize that we also include the multiplicative errors $\Delta\phi_j$ in the processing phases $\phi_j$ due the limitations of the arbitrary waveform generator \cite{werninghaus2021leakage,zurichinst}.

\subsection{Simple scheme to reduce the unwanted effects of always-on couplings}

Here, we briefly discuss the cancellation schemes that we use to implement the QSP-based generalized parity measurement on the superconducting cavity QED platform \cite{blais2021circuit}. Notice that going to the interaction picture with respect to the dressed cavity and qubit Hamiltonians in Eq. ($\ref{eq:FullHamiltonian}$), we not only obtain $H_{\chi}\equiv \chi\sigma_z\otimes n_C$, but also to terms of second order in $n_C$ 
\begin{align}
H_{\bar{K}} \equiv \frac{\bar{K}}{2} \sigma_z \otimes n_C^2 \quad\quad H_{K} \equiv  \frac{K_C}{2} n_C^2,
\end{align}
where $\bar{K}\equiv K_C\eta$. 

Both nonlinear terms are unwanted if we want to implement the QSP protocol with high fidelity. Moreover, the always-on nature of $H_{\chi}$ makes it difficult to implement high-fidelity single-qubit processing unitaries, especially in the presence of a large number of cavity photons [ note that the single qubit Rabi frequency cannot be increased arbitrary without jeopardizing the two-level approximation in Eq. ($\ref{eq:JCHam}$)]. Here, we briefly discuss the strategies to reduce the effects of these unwanted evolutions for the state preparation protocol.     

First, we emphasize that the cavity Kerr nonlinearity represented by $H_{K}$ commutes with all operations that make up the QSP protocol $\tilde{U}^{\mathbf{R}}_{\vec{\Phi}}(n_C,r,k)$ (see main text). Hence, it is possible to compensate for the effect of $H_{K}$ at the end of the QSP protocol. This compensation can be implemented either in post-processing, or by implementing a strong Kerr non-linearity with the opposite sign to $K$ by driving the measurement qubit near its transition frequency (see Ref. \cite{zhang2022drive} for the details of engineering drive-induced nonlinearities in superconducting cavity QED).

In contrast, the always-on linear and nonlinear qubit-cavity couplings result in unwanted coherent dynamics that do not commute with the QSP-based measurement protocol. Still, as we argue below, it is possible to strongly reduce the effect of this unwanted evolution at will by only controlling the measurement qubit. To this end, notice that the Hamiltonians $H_{\chi}$ and $H_{\bar{K}}$, describing the linear and nonlinear coupling between the measurement qubit and the cavity have the product form, (i.e., such that the eigenvectors of both $H_{\chi}$ and $H_{\bar{K}}$ are separable). This property allows us to reduce the effect of $H_{\chi}$ and $H_{\bar{K}}$ at will, using only $\sigma_z$ rotations on the measurement qubit, given that the photon number variance of the cavity state is sufficiently small. In particular, the evolution due to the average photon number can be completely cancelled through the following replacements
\begin{align}
e^{-itH_{\bar{K}}} \leftarrow e^{-it(H_{\bar{K}} - \frac{\bar{K}}{2}\bar{n}^2\sigma_z)}\quad \quad e^{-itH_{\chi}} \leftarrow e^{-it(H_{\chi} - \chi\bar{n}\sigma_z)},
\label{eq:Replacements}
\end{align}
where $\bar{n}$ is the average photon number of the cavity state and $t$ is the time of evolution. 

In our numerical simulations, we reduce the effect of $H_{\bar{K}}$ throughout the implementation of the state-preparation protocol, while the replacement for $H_{\chi}$ is only used during the implementation of the processing unitaries. We note that during the implementation of the signal unitaries, the corresponding replacement in Eq. ($\ref{eq:Replacements}$) can be implemented efficiently using virtual $Z$ gates by changing the phase of the resonant drives implementing the processing unitaries \cite{mckay2017efficient}.

\begin{table}[]
    \centering
    \begin{tabular}{c|c || c| c}
         $\Delta_{\phi}/\phi$ & 0.01 & $\bar{T}_c$ & 25 ms \\ \hline
         $K_q$ &  $2\pi\times 246$ MHz &   $\gamma_{c}=\bar{T}_c^{-1}$ &  $2\pi\times 3.2$ Hz  \\  \hline
         $\Omega_q$ & $2\pi\times 8.2$ MHz &   $\gamma_{q,\phi}/ \gamma_{q}$ & 4.5 \\ \hline
         $\chi$ & $2\pi\times 41$ kHz &  $\bar{T}_{1}$ & 300 $\mu$s \\ \hline
         $n^e_{\rm crit} $ & 1000 & $\gamma_{q}= \bar{T}_1^{-1}$&  $2\pi\times 0.53$ kHz  \\ \hline
         $K_C$ & $2\pi \times 2$ Hz& $\Delta = 9E_C$ & $2\pi \times 2.46$ GHz \\ \hline
    \end{tabular}
    \caption{ The parameters describing the cavity QED setup considered for the state preparation protocols. The relaxation times $\bar{T}_{c}$ and $\bar{T}_1$ denotes the quantities relevant to dispersive regime.}
    \label{tab:gatecount2}
\end{table}

\section{State Preparation Protocols}

Here, we provide the details for preparing large $r$-component cat states using sequences of generalized parity measurements as discussed in the main text. We also consider the effect of decoherence on the fidelity of the state preparation protocols for the $r$-component cat state preparation. 

We begin by detailing the pulse sequences that are used in the numerical simulations of the state preparation protocols. We represent the operations that constitute each state preparation as a sequence of unitary operations and projective measurements of the measurement qubit on the computational basis, denoted $``{\rm--}"$. In the sequences below each operation is followed by the operation to its left.

The $r$-component Cat state preparation can be symbolically represented as 
\begin{align}
{\rm CAT}_{r,k;1} \equiv {\rm--} {\rm H}\tilde{U}_{\vec{\Phi}}(H_S,r,k){\rm H}\equiv  {\rm--} \tilde{U}^{\rm CQED}_{\vec{\Phi},r,k},
\end{align}
 where $H_S\equiv H_{\rm QND}$ [see Eq. (\ref{eq:FullHamiltonian})] and
 \begin{align}
\tilde{U}_{\vec{\Phi}}(H_S,r,k) =  e^{i \Phi_0 Z}e^{iX \otimes H_{S,r,k} }e^{i \Phi_1 Z}e^{iX \otimes H_{S,r,k} X}\cdots e^{iX \otimes H_{S,r,k}}e^{i \Phi_d Z}
 \end{align}
 where $H_{S,r,k} \equiv \frac{\pi}{r} ( H_{S}/\chi - k)$ and $\tilde{U}^{\rm CQED}_{\vec{\Phi}}$ is a QSP protocol where the single-qubit operators $X$ and $Z$ acting on the measurement qubit are swapped. We prefer this notation to describe our similations because the Hamiltonian describing the native qubit-cavity interaction in cavity QED setups is $H_{\chi}\equiv \chi \sigma_z \otimes a^{\dagger}a$. 
  Notice that in ${\rm CAT}_{r,k;1}$ we removed the final $X$ operation on the measurement qubit compared to the protocol discussed in the main text. Hence, for the above protocol, the state preparation is successful when the measurement qubit is found in $\ket{1}$.

As mentioned earlier, the QNDness of the measurement protocol can be leveraged to improve fidelity of the prepared state through repeated measurement, as long as the errors that do not respect the QNDness do not dominate (i.e., errors induced by the dephasing of the measurement qubit). 
In the case of $s$ repeated measurements, we simply repeat a single protocol $s$ times 
\begin{align}
{\rm CAT}_{r,k;s}\equiv \left[ {\rm--}\tilde{U}^{\rm CQED}_{\vec{\Phi},r,k}\right]^s .
\label{eq:CATrepeat}
\end{align}
Then the state preparation is successful only if the $s$ projective measurements find the measurement qubit alternating between $\ket{1}$ and $\ket{0}$.

\subsection{Success probability of the perfect state preparation protocol in the absence of dissipation}

\begin{figure}
    \centering
    \includegraphics[width=0.45\textwidth]{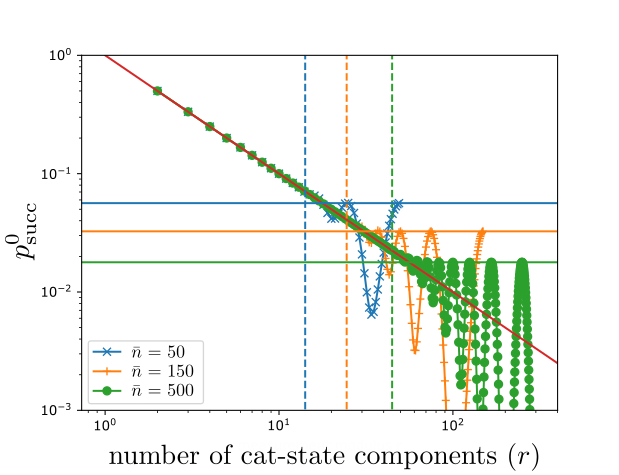}
    \caption{A log-log plot demonstrating the scaling of the success probability $p_{\rm succ}^0$ for preparing $r$-component cat states with $\bar{n}\in \{50,150,500\}$ photons as a function of $r$, (in blue, yellow, and green, respectively). For $r\leq 2\sqrt{\bar{n}}$, the $p_{\rm succ}\approx 1/r$ while for $r> 2\sqrt{\bar{n}}$ we observe oscillations whose maximum is $|\langle \alpha=\sqrt{\bar{n}}\ket{\bar{n}}|^2=O(1/\sqrt{\bar{n}})$. }
    \label{fig:succ_probs}
\end{figure}

Before we continue with our perturbative analysis of the effects of decoherence on the proposed state preparation protocols, we briefly present the success probabilities for the perfect state preparation protocols starting from different coherent states as a function of the modulus $r$. The success probability $p^0_{\rm succ}$ is given by the probability of measuring the ancillary measurement qubit in the $\ket{1}$ state given that the initial state is $\ket{0}\ket{\alpha=\sqrt{\bar{n}}}$, where the state of the second register is a coherent state with an average photon number of $\bar{n}$. In the absence of decoherence, the success probability is simply given by the square of the overlap between $\ket{\alpha=\sqrt{\bar{n}}}$ and the normalized desired state $\ket{\phi^{\bar{n}}_{\rm des}}\equiv G_r\ket{\alpha}/||G_r\ket{\alpha}||$, where $G_r\equiv \Pi_{S_{r,0}}$ is an orthogonal projector on to the $S_{r,0}$ that is spanned by states $\{\ket{m}\}$ with $m=0 \,{\rm mod}\, r$ excitations. The results for the success probability for $\bar{n}\in \{50,150,500\}$ (in blue, yellow, and green, respectively) as a function of $r$ is shown in Fig. $\ref{fig:succ_probs}$. As discussed in the main text, $p^0_{\rm succ}\approx 1/r$ for $r\leq 2\sqrt{\bar{n}}$ (indicated by vertical dashed lines with matching colors). On the other hand, for $r> 2\sqrt{\bar{n}}$ we observe oscillations in the success probability, as expected from the Gaussian distribution of different number states around $\bar{n}$. The peak success probability of these oscillations correspond to the squared overlap between a coherent state $\ket{\alpha=\sqrt{\bar{n}} }$ with average photon number $\bar{n}$ and a photon number state $\ket{\bar{n}}$ with $\bar{n}$ photons (indicated by the horizontal lines with matching colors). 

\section{Effect of various decoherence channels} 
To study the effects of decoherence on the state preparation protocols, we develop a methodology that is i) inspired by the quantum trajectories description of open quantum systems \cite{castin2008wave,dalibard1992wave} and ii) is enabled by the single-qubit description of the QSP protocols. In particular, we use the single-qubit description of the QSP protocol where the signal unitary is given by $e^{i\theta_{m}^{r,0} X}$, with $\theta_m^{r,0} = \frac{\pi}{r} \times m$ (corresponding to the measured cavity having $m$ photons), to estimate the fidelity of the post-measurement state given that the protocol succeeds \textit{despite} the occurrence of one of the errors described in the previous section.

\subsubsection{The framework to study the effects of decoherence on QSP protocols}
Our approach starts with the stochastic description of the wavefunction evolution as first described by \cite{dalibard1992wave,castin2008wave}. In this framework, the evolution of an initial pure wavefunction $\ket{\phi(t)}$ between times $t$ and $t+\Delta t$ for small enough $\Delta t$ is described by either one of the following processes
\begin{enumerate}
    \item No quantum jump occurs with probability $1-\delta p$ and the new state is given by 
    \begin{align}
    \ket{\phi(t+\delta t)} = \frac{1}{\sqrt{\mathcal{N}_{\rm nj}}}\exp\left(-i\bar{H}(t)\delta t\right) \ket{\phi(t)}, 
    \label{eq:nojumpstate}
    \end{align}
    where $\mathcal{N}_{\rm nj}$ is the appropriate normalization factor. Here $\bar{H}$ is non-Hermitian generator of evolution in the absence of a quantum jump, given by 
    \begin{align}
    \nonumber &\bar{H}
    \equiv H + H_{\rm NH} \\
    &H_{\rm NH} \equiv - \frac{i}{2}\sum_{m}L_{m}^{\dagger}L_m,
    \label{eq:nojumpham}
    \end{align}
    which is a combination of the Hermitian generator of unitary dynamics $H$ and the non-Hermitian component $H_{\rm NH}$ deriving from the jump operators $\{L_{m}\}$ similar to those defined in the previous section. Up to first order in $||\delta t H_{\rm NH}||$, evolution occurs without quantum jumps with probability $\mathcal{N}_{\rm nj} \approx 1 - \delta t \sum_{m}\bra{\phi(t)}L_{m}^{\dagger}L_m \ket{\phi(t)} \equiv 1-\sum_m \delta p_m = 1-\delta p$.
    \item A quantum jump $L_m$ occurs, transitioning $\ket{\phi(t)}$ to 
    \begin{align}
    \ket{\phi(t+\delta t)} = \frac{L_m\ket{\phi(t)}}{\sqrt{\mathcal{N}_{m}}},
    \label{eq:jumpstate}
    \end{align}
    with the appropriate normalization factor $\mathcal{N}_{m} = \bra{\phi(t)}L_m^{\dagger}L_m\ket{\phi(t)}$, and the probability for a quantum jump to occur is $ \delta p_m\equiv \delta t \mathcal{N}_{m}$. 
\end{enumerate}

The above equations are valid when the Hermitian generator is not time-dependent. In order to leverage the above description to study the effect of errors for our state-preparation protocol, we propagate these equations through a piece-wise time-dependent Hamiltonian $H^{(r)}_{\rm QSP}(t)$, which describes the proposed implementation of the QSP-based state preparation in the absence of any decoherence channels. Following the processing angles in Eq. ($\ref{eq:AngleFormula}$), each QSP protocol consists of $\tilde{r}=2\times\ceil{r/2}-1$ iterations of the signal operator. 
In the following, $\{\delta t_l\}$ to denote the time duration of each iteration of the QSP protocol and $T \approx \sum^{\tilde{r}}_l \delta t_l$ to denote the total duration of the state preparation protocol. The unitary evolution operator that describe the QSP protocol up to the end of the $l^{\rm th}$ iteration is denoted as $U_{{\rm QSP},l}$. We also use the notation $U_{{\rm QSP},\tilde{r}}\equiv U_{\rm QSP}$ for simplicity. 

We approximate the dissipative evolution during the state preparation protocol by only considering resulting the wavefunction  up to first order in $\delta p_m$. Moreover, we simplify the treatment of the quantum jumps by assuming that the jump events can only happen right after the application of a signal unitary. For this analysis, we assume that the signal unitary is generated only by $H_{\chi} = \chi \sigma_z \otimes a^{\dagger}a$, and ignore the nonlinear coupling $H_{\bar{K}}$. 

Propagating the initial state $\ket{\phi_{\rm init}}$ using the above stochastic equations in Eqs. ($\ref{eq:nojumpstate}$),($\ref{eq:nojumpham}$), and ($\ref{eq:jumpstate}$) under the restriction that only a single quantum jump can occur, we find that the final state $\ket{\phi_{\rm final}}$ is given by one of the following:
\begin{enumerate}
\item{\textit{In the case of no quantum jumps:}} The final quantum state is 
\begin{align}
\ket{\phi_{\rm final}^{\rm nj}} = \frac{1}{\sqrt{M_{\rm nj}}}\mathcal{T}\int_{0}^{T} \exp{-i\bar{H}_{\rm QSP}(t) }dt \ket{\phi_{\rm init}},
\end{align}
where $\mathcal{T}$ is the time ordering operator, $T$ is the total time to implement the state preparation protocol, and $\bar{H}_{\rm QSP}(t) \equiv H_{\rm QSP}(t) + H_{\rm NH}$. The normalization constant $\mathcal{M}_{\rm nj}$ gives the probability that no quantum jumps occur.

By expanding the time-ordered evolution operator up to first order in $H_{\rm NH}$, we obtain the approximation
\begin{align}
\ket{\phi_{\rm final}^{\rm nj}}\approx \frac{1}{\sqrt{M_{\rm nj}}} U_{\rm QSP}\left(1 - iT\bar{H}_{\rm NH}\right)\ket{\phi_{\rm init}}.
\end{align}
Crucially, the above expression involved a new effective non-Hermitian Hamiltonian $\bar{H}_{\rm NH}$ which describes the effect of dissipation in the absence of quantum jumps while taking into account the time-dependent evolution that implements the QSP protocol. The effective Hamiltonian can be expanded as
\begin{align}
\bar{H}_{\rm NH} = -\frac{i}{2}\sum_{m}\sum_{l=1}^{\tilde{r}}\frac{\delta t_l}{T} \bar{L}_{m,l}^{\dagger}\bar{L}_{m,l},
\end{align}
where
\begin{align}
\bar{L}_{m,l}\equiv \left(\Pi_{k=l+1}^{\tilde{r}}U_{{\rm QSP},k}\right)L_m\left(\Pi_{k'=1}^{l}U_{{\rm QSP},k'}\right),
\end{align}
describe the effect of a jump operation that occurs between the $l^{\rm th}$ and $l+1^{\rm st}$ iterations of the signal unitary. The probability that no such jump operator occurs during the state preparation protocol is 
\begin{align}
 \mathcal{N}_{\rm nj}\approx 1-\sum_{m}\sum_{l=1}^{\tilde{r}} \delta t_l ||\bar{L}_{m,l}^{\dagger}\ket{\phi_{\rm init}}||^2 \equiv 1- \sum_{m}\sum_{l=1}^{\tilde{r}}\bar{\delta} p_{m,l}.
\end{align}

\item \textit{The case where initial state has gone through a quantum jump:} The final quantum state is described by the modified jump operators $\bar{L}_{m,l}$
\begin{align}
\ket{\phi_{m,l}} = \frac{1}{\sqrt{\mathcal{M}_{m,l}}} \bar{L}_{m,l}\ket{\phi_{\rm init}},
\end{align}
with probability $\bar{\delta} p_{m,l}$.
\end{enumerate}

Using the definition of a density matrix \cite{nielsen2002quantum}, it is possible to express the state resulting from the constrained stochastic evolution described above conveniently as
\begin{align}
 \sigma_{{\rm final},r} =  U_{\rm QSP}\left[\sigma_{\rm init} - i T \{\bar{H}_{\rm NH},\sigma_{\rm init}\} \right] U_{\rm QSP}^{\dagger}
+ \sum_{m}\sum_{l}\delta t_{l} \bar{L}_{l,m}\sigma_{\rm init}\bar{L}_{l,m}^{\dagger}.
\end{align}
Next, we use the expression for $\sigma_{\rm final}$ to calculate the success probability and the post-measurement state fidelity associated with the state preparation protocol.

\subsubsection{The calculation of the state preparation Fidelity for a cavity QED system}


Next, we calculate the success probability and the fidelity of $\sigma_{\rm final}$ with respect to a desired state $\sigma_{\rm des}\equiv\ket{1}\bra{1}\otimes\ket{\phi_{\rm des}}\bra{\phi_{\rm des}}$, assuming that the desired state is  $\ket{\phi_{\rm des}}=\ket{1}\left(G_r\ket{\alpha}/\sqrt{\mathcal{N}_r}\right) \equiv \ket{1}\ket{G_r}$ with $\mathcal{N}_r\equiv ||G_r\ket{\alpha}||^2$ (we remind the reader that $G_r\equiv \Pi_{S_{r,0}}$). Then the fidelity and the success probability are given by 
\begin{align}
p_{\rm succ}\equiv \mathrm{Tr}[\sigma_{\rm final } \Pi_1]\quad \mathrm{and}\quad \mathcal{F}\equiv\sqrt{\mathrm{Tr}[\Pi_1\sigma_{\rm final } \Pi_1 \sigma_{\rm des}]/p_{\rm succ}},
\end{align}
respectively (here, we used $\Pi_{\sigma} \equiv \ket{\sigma}\bra{\sigma}\otimes \mathbf{I}$ as in the main text) . The success probability given an initial pure state $\sigma_{\rm init}\equiv \ket{\phi_{\rm init}}\bra{\phi_{\rm init}}$ where $\ket{\phi_{\rm init}}\equiv \ket{0}\ket{\phi_{\rm init}^c}$ can be calculated as 
\begin{align}
\nonumber p_{\rm succ} = ||G_r\ket{\phi^c_{\rm init}}||^2 + \sum_{m,l}\delta t_l|| \Pi_1 \bar{L}_{m,l}\Pi_0\ket{\phi_{\rm init}}||^2 -iT{\rm Tr}\left[\ket{0}\bra{0}\otimes G_r\left\{ \bar{H}^{00}_{\rm NL},\sigma_{\rm init}\right\}\right],
\end{align}
where we define $\bar{H}^{00}_{\rm NL} = \Pi_0\bar{H}_{\rm NL}\Pi_0$. To derive the last term, we used the the fact that $G_r^2=\mathbf{1}$, by the virtue of being an orthogonal projector. 
This term can be further simplified if $\ket{\phi_{\rm init}}=\ket{0}\ket{\alpha}$
to obtain 
\begin{align}
p_{\rm succ} = \mathcal{N}_r + \sum_{m,l}\delta t_l|| \Pi_1 \bar{L}_{m,l}\Pi_0\ket{\phi_{\rm init}}||^2 - 2i T \sqrt{\mathcal{N}_r}\mathrm{Re}\left[\bra{\alpha}\tilde{H}^{00}_{\rm NH}\ket{G_r}\right],
\end{align}
where $\tilde{H}^{00}_{\rm NH}\equiv (\bra{0}\otimes \mathbf{I} )\bar{H}^{00}_{\rm NH} (\ket{0}\otimes\mathbf{I})$.

Next, we calculate the matrix elements of $\tilde{H}^{00}_{\rm NH}$ for the three decoherence channels that are most relevant for an implementation on the superconducting cavity QED platform. The bare jump operators $\{L_m\}$ for these channels are given in Eq. ($\ref{eq:qubitDeph}$). Fortunately, the $\bar{H}_{\rm NH}$ for qubit dephasing and cavity decay channels are very straightforward to calculate because operators $Z^2$ and $a^{\dagger}a$ both commute with $U_{{\rm QSP},l}$ for any $l$. That is, 
\begin{align}
\bar{L}^{\dagger}_{cr,l}\bar{L}_{cr,l} = \gamma_c \mathbf{I}_{\mathcal{Q}}\otimes \hat{n}_C \quad\quad
\bar{L}^{\dagger}_{qz,l}\bar{L}_{qz,l} = \frac{\gamma_{qz}}{2} \mathbf{I},
\end{align}
where $\mathbf{I}_{\mathcal{Q}}$ and $\mathbf{I}$ are identity operators acting on the measurement qubit and the whole system, respectively. Finally, the relevant expectation values for the qubit decay channel is given by 
\begin{align}
\nonumber \Pi_0\bar{L}^{\dagger}_{q,r}\bar{L}_{q,r}\Pi_0 &=\gamma_{q,r}\Pi_0U_{{\rm QSP},l}\Pi_{1} U_{{\rm QSP},l}^{\dagger}\Pi_0.
\label{eq:qr_opt}
\end{align}
Conveniently, these matrix elements can be easily calculated using a simulation of the relevant parts of the QSP unitary based on a single qubit. In particular, we can write the matrix elements of the operator in Eq. ($\ref{eq:qr_opt}$) as 
\begin{align}
\nonumber \bra{\phi_{\rm init}}\Pi_0\bar{L}^{\dagger}_{q,r}\bar{L}_{q,r}\Pi_0 \ket{G_r}&= \frac{\bra{\phi_{\rm init}}\Pi_0\bar{L}^{\dagger}_{q,r}\bar{L}_{q,r}\Pi_0 G_r \ket{\phi_{\rm init}}}{\sqrt{\mathcal{N}_r}}\\
&=\frac{\sum_{n\in S_{r,0}}|\alpha_n|^2 |\bra{0}U_{{\rm QSP},l}(n)\ket{1}|^2}{\sqrt{\mathcal{N}_r }}.
\end{align}
In the last step above, we defined $U_{{\rm QSP},l}(n)$ as a unitary acting only on the measurement qubit, which satisfies $U_{{\rm QSP},l}(\ket{0}\otimes\ket{n}) = (U_{{\rm QSP},l}(n)\ket{0})\otimes\ket{n}$. 


Similarly, the fidelity squared can then be approximated as
\begin{align}
\nonumber \mathcal{F}^2 \approx &\frac{\mathcal{N}_r -i T\sqrt{\mathcal{N}_r}\mathrm{Re}\left[\bra{\alpha}\tilde{H}^{00}_{\rm NH}\ket{G_r}\right] +\sum_{m,l}\delta t_l |\bra{\phi_{\rm des}}\bar{L}_{m,l}^{10}\ket{\phi_{\rm init}}|^2}{p_{\rm succ}} \\
\nonumber \approx &1 - \frac{\sum_{m,l}\delta t_l\left(||\bar{L}^{10}_{m,l}\ket{\phi_{\rm init}}||^2 - ||\bra{\phi_{\rm des}}\bar{L}^{10}_{m,l}\ket{\phi_{\rm init}}||^2\right)}{\mathcal{N}_r}\\
=& 1- \frac{\sum_{m,l}\delta t_l \left|\left|\Pi_{\perp_{\ket{\phi_{\rm des}}}}\bar{L}^{10}_{ml} \ket{\phi_{\rm init}}\right|\right|^2}{\mathcal{N}_r},
\label{eq:Pert_fidelity}
\end{align}
where in the second approximation, we plugged in the definition of $p_{\rm succ}$ and kept only terms to the lowest order in $\delta p_m$, and in the final equality we simply defined the orthogonal projector onto the complement of $\ket{G_r}$
\begin{align}
\Pi_{\perp_{\ket{\phi_{\rm des}}}}\equiv \mathbf{I} - \ket{\phi_{\rm des}}\bra{\phi_{\rm des}}.
\end{align}

\subsubsection{Numerical Comparisons}

Here we provide evidence that the perturbative framework we introduced above to approximate the effect of decoherence for QSP protocols agree well with the numerical simulations of the full master equation describing the dynamics of the cavity QED system. We emphasize again that for the purposes of this section, we do not consider the cavity self-Kerr term and nonlinear coupling between the qubit and the cavity in Eq. ($\ref{eq:FullHamiltonian}$) 

\begin{figure}[h!]
    \centering
    \includegraphics[width=0.45\textwidth]{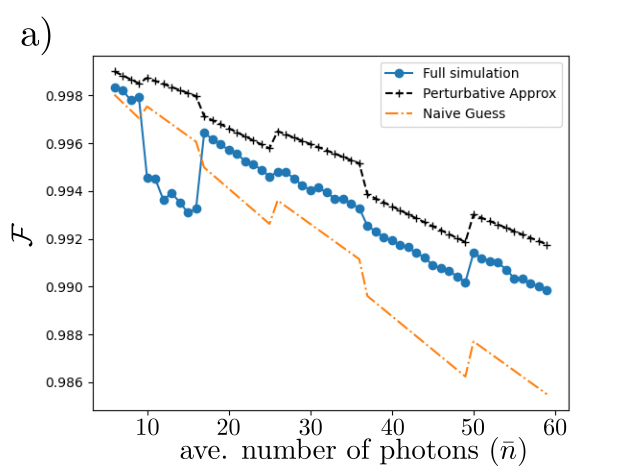}
    \includegraphics[width=0.45\textwidth]{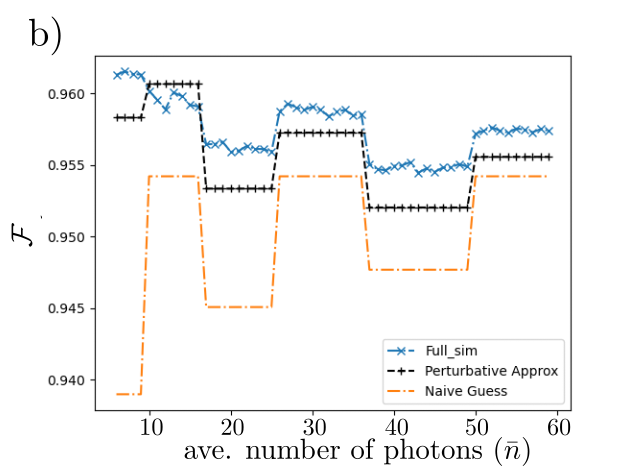}
    \includegraphics[width=0.45\textwidth]{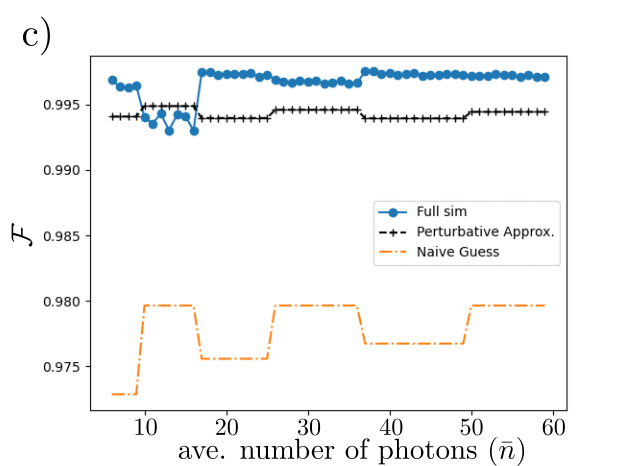}
    \includegraphics[width=0.45\textwidth]{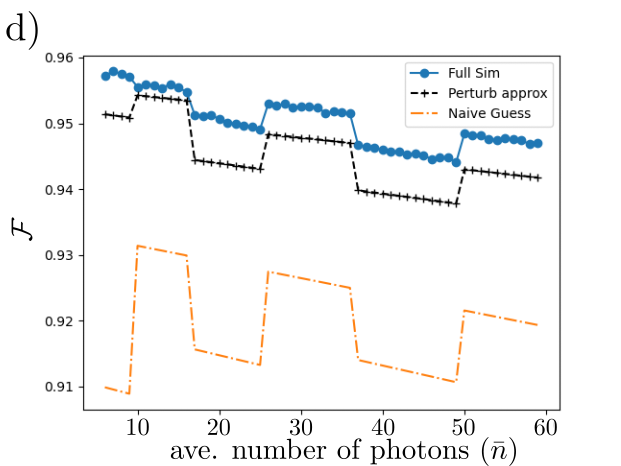}
    \caption{The comparison between the values of the fidelity, obtained by i) the simulation of the full master equation ii) evaluation of the perturbative expression in Eq. ($\ref{eq:Pert_fidelity}$) and the naive expression for the fidelity in Eq. ($\ref{eq:Naive_fidelity}$) , for the three most relevant error mechanisms in the setting of cavity QED. For all results in this section, we ignore the effects of the nonlinear coupling between the cavity and the measurement qubit. a) The comparison of fidelities given that the only decoherence channel is given by the decay of the cavity photons. Remarkably, the cavity decay rate is effectively reduced by half. b) The same as a) but assuming that the only decoherence mechanism is the dephasing of the measurement qubit. The dephasing of the measurement qubit is remains the most important contributor to the reduction of fidelity. c) The same as a) but assuming that the only decoherence mechanism is the decay of the measurement qubit. The contribution of the qubit decay events on the overall infidelity is greatly reduced. d) The same as a) but when all three decoherence mechanisms are taken into account. }     \label{fig:PerturbComparison}
\end{figure}

The comparison of fidelity for the three decoherence channels most relevant to the cavity QED platform are given in Fig. $\ref{fig:PerturbComparison}$, for the task of preparing a $\ceil{\sqrt{\bar{n}}}$-component cat state with a $\bar{n}$ average photon number. We compare the results from the full simulation of the dissipative dynamics and the  perturbative treatment with what we expect from a naive guess, which can be simply expressed as (dot-dashed)
\begin{align}
\mathcal{F}_{\rm naive}(\bar{n}) \approx 1- \frac{T_{\ceil{\sqrt{\bar{n}}}}}{2}\left(\gamma_{cr}\bar{n} + \gamma_{qr} + \frac{\gamma_{qz}}{2} \right),
\label{eq:Naive_fidelity}
\end{align}
where $T_{\ceil{\sqrt{\bar{n}}}}$ is the total time to implement the QSP based state preparation protocol.
We observe that our perturbative approximation improves the naive guess for each one of the decoherence channels. We also note that a similar improvement can be observed for the success probability (figure not shown). 

\begin{figure}[h!]
    \centering
    \includegraphics[width=0.45\textwidth]{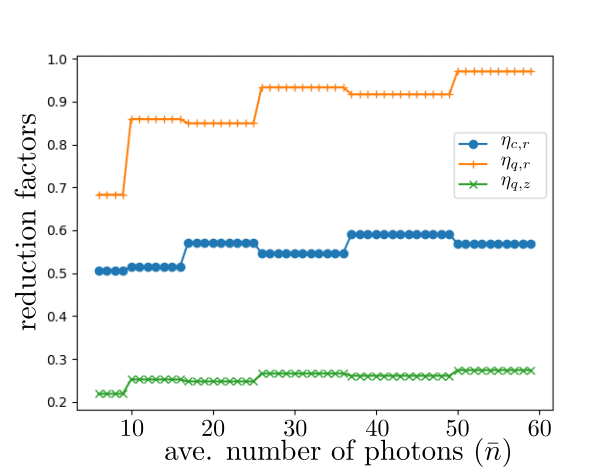}

    \caption{The correction factors [see Eq. ($\ref{eq:correction_factors}$)]obtained using the first-order perturbation theory discussed above as a function of the average photon number $\bar{n}$ for the $\ceil{\sqrt{\bar{n}}}$-component cat state preparation.}    \label{fig:correction_factors}
\end{figure}

The perturbative result can be expressed in a similar form as $\mathcal{F}_{\rm naive}$ using $\bar{n}$-dependent correction factors for each decoherence channel
\begin{align}
\mathcal{F}_{\rm pert}(\bar{n}) = 1- \frac{T_{\ceil{\sqrt{\bar{n}}}}}{2}\left(\eta_{cr}(\bar{n})\gamma_{cr}\bar{n} + \eta_{qr}(\bar{n})\gamma_{qr} + \eta_{qz}(\bar{n})\frac{\gamma_{qz}}{2} \right).
\label{eq:correction_factors}
\end{align}
To highlight the robustness of the QSP based cat state preparation protocol to irreversible errors, we plot the correction factors for $\bar{n}<60$ in Fig. $\ref{fig:correction_factors}$. Our results show that the QSP-based state preparation protocol reduced the adverse effects of cavity decay on fidelity by $\approx 1/2$ and those due to qubit decay is reduced by $\approx 1/4$, while the effect of of qubit dephasing is similar to its bare counterpart. Notice that the abrupt changes in Figs. $\ref{fig:PerturbComparison}$ and $\ref{fig:correction_factors}$ are due to those in $T_{\ceil{\sqrt{\bar{n}}}}$ as we change the modulus of the generalized parity measurement between an odd and even number (see Eq. $\ref{eq:AngleFormula}$).

Finally, we emphasize that the discussion in this section focused on a single application of the QSP-based cat state preparation starting from an initial coherent state. Yet, because the effective measurement that we implement through the QSP protocol is a QND measurement, repeating the measurement can easily improve the fidelity of the post-measurement state. The errors are further reduced due to the fact that the initial cavity state for the second round of measurement is already close to the desired cat state.


\section{Dependence of $r$-component cat state preparation fidelity on $r$ for $\bar{n}\in \{50,150,300,378,500\}$. }

\begin{figure}[h!]
    \centering
    \includegraphics[width = 0.45\textwidth]{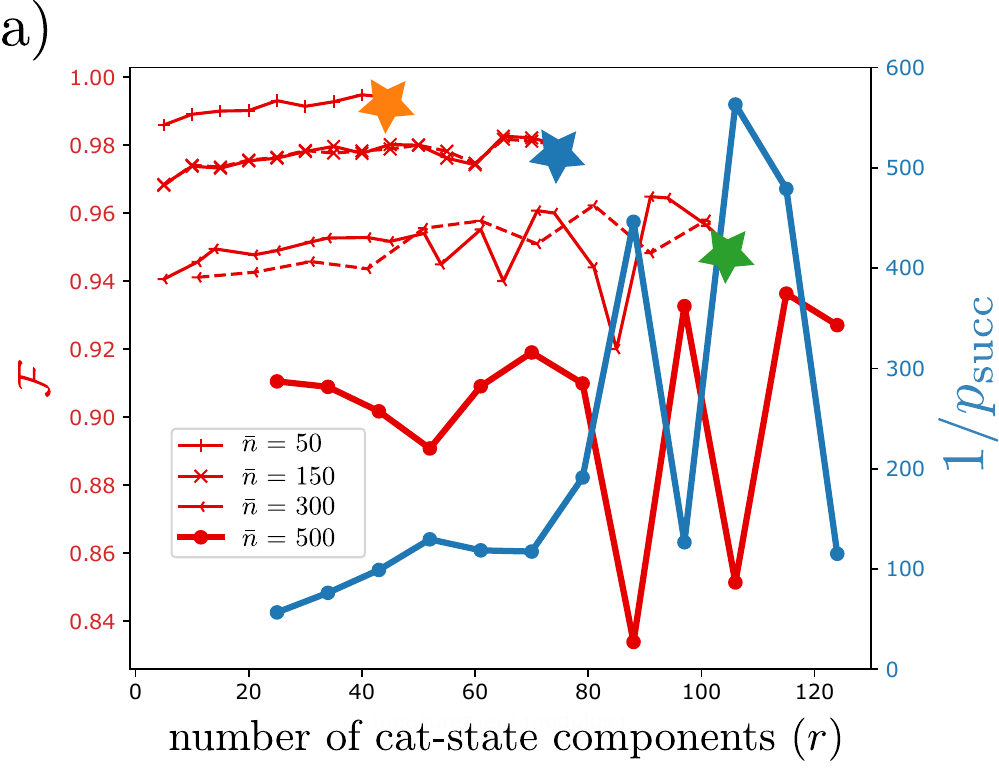}
    \hspace{10mm} \includegraphics[width = 0.45\textwidth]{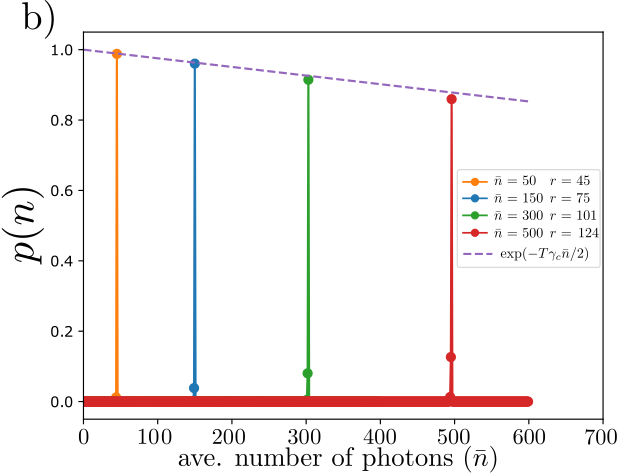}
    \includegraphics[width = 0.45\textwidth]{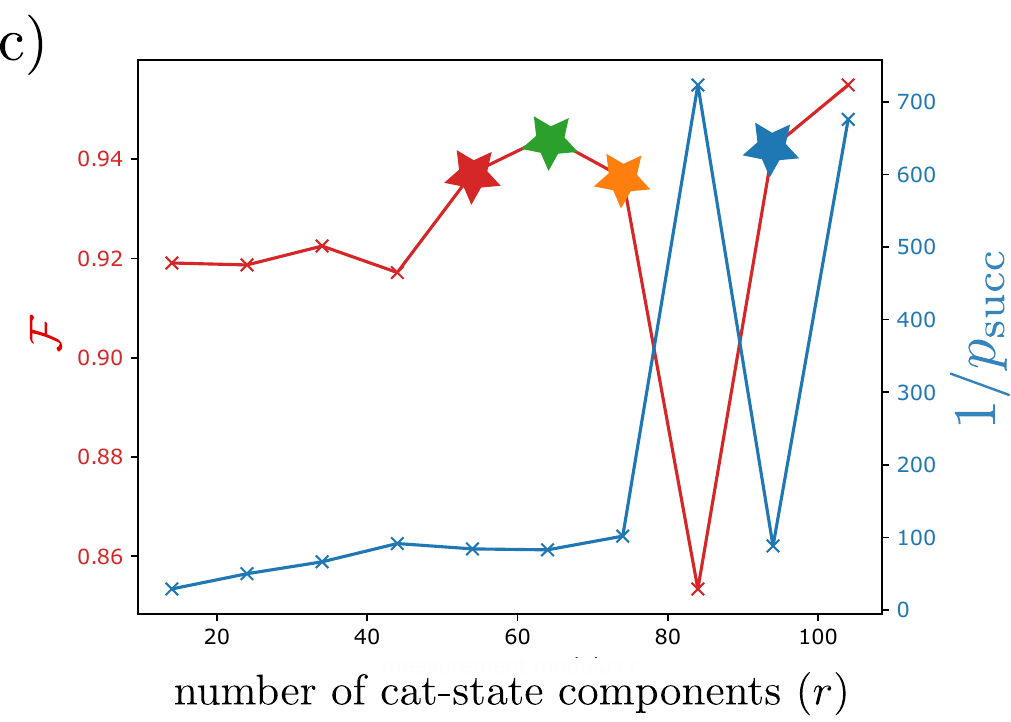}
    \hspace{10mm}
    \includegraphics[width = 0.45\textwidth]{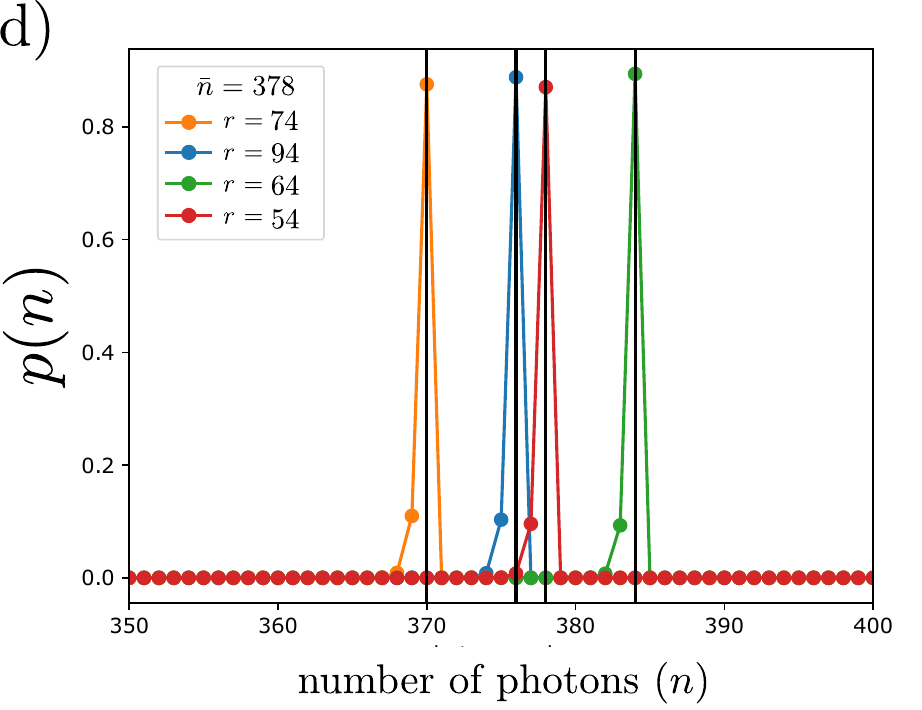}

    \caption{The $r$-dependence of the state preparation protocol  CAT$_{r,0;3}$ [see Eq. ($\ref{eq:CATrepeat}$)]. a) The dependence of the fidelity as a function of $r$ for average photon numbers $\bar{n}\in \{50,150,300,500\}$. Aside from a slight increase, the state preparation fidelities do not significantly change as a function of $r<2\sqrt{\bar{n}}$ for $\bar{n}\leq 300$, both for an implementation with $H_{\bar{K}}=0$  (full lines) and $H_{\bar{K}}\neq 0$ (dashed lines). The oscillations for $r>2\sqrt{\bar{n}}$ correlate with reduced success probability, which is plotted for the case of $\bar{n}=500$ for illustration (thick blue trace). The peak values of the fidelity oscillations correspond to photon number states with $\approx \bar{n}$ photons. We show this in b), by plotting the number distributions of the prepared states for $r$ values indicated by a star in a). The highest photon number occupation of the prepared states matches that of a Fock state evolving in a cavity whose decay rate is $\gamma_c/2$ for a time $T_{r}$. c) The fidelity and inverse success probability of the state preparation protocol for $\bar{n}=378$ and for various $r$, with $H_{\bar{K}}\neq 0$. d) We find that the post-measurement state approximates a Fock state when $r\in\{54,64,74,94\}$ [marked by stars in c)], with associated fidelity $\approx 94\%$. }
    \label{fig:r_dep_data}
\end{figure}

In this last subsection, we present numerical results demonstrating that the fidelity of the state conditioned on the success of the state preparation protocol CAT$_{r,0;3}$ does not depend strongly on the modulus $r$. In Fig. $\ref{fig:r_dep_data}$ a) we present the numerical results describing the dependence of the fidelity on $r$ for $\bar{n}\in \{50,150,300,500\}$.
Here, the results of simulations with the nonlinear coupling Hamiltonian is switched off ( $H_{\bar{K}}=0$) and switched on ($H_{\bar{K}}\neq 0$) are indicated using full and dashed lines, respectively. For $r>\ceil{\sqrt{\bar{n}}}$ small oscillations appear in the fidelity. As shown in Fig. $\ref{fig:r_dep_data}$ a) for the case of $\bar{n}=500$ with $H_{\bar{K}}=0$ (thick traces), the fidelity is decreased for the values of $r$ where the success probability becomes very small. We emphasize that the fidelity of the state preparation does not change significantly depending on whether $H_{\bar{K}}=0$ or $H_{\bar{K}}\neq 0$, for $\bar{n}\in\{50,150,300\}$

For all cases considered in Fig. $\ref{fig:r_dep_data}$ a), the largest values of the modulus $r$ are large enough such that the post-measurement state well approximates a photon-number state. For completeness, in Fig. $\ref{fig:r_dep_data} $ b), we plot the photon number distributions $p(n)$ of the outputs of CAT$_{r,0;3}$ for the largest $r$ plotted in Fig. $\ref{fig:r_dep_data}$ a) (indicated by a star for each $\bar{n}$). The overlap of the prepared states with a photon number state can be approximated by only considering the loss of fidelity of a photon number state evolving under a cavity whose decay rate is $\gamma_c/2$ (indicated by the dashed line). 

Finally, we focus on the $r$-dependence of fidelity and success probability in the case of $\bar{n}=378$ using a simulation of CAT$_{r,0;3}$ with $H_{\bar{K}}\neq 0$. In Fig. $\ref{fig:r_dep_data}$ c), we demonstrate that the fidelity does not change significantly with increasing $r$, while the inverse success probability increases as expected. For $r\in\{54,64,74,94\}$ (indicated by stars), the post-measurement states approximate photon number states, and the resulting fidelities are higher than those with lower $r$. We depict the photon number distributions of the post-measurement states in  $\ref{fig:r_dep_data}$ d). The peaks of the distributions coincide with integer multiples of the modulus $r$ (black vertical lines).  

